\documentclass[Times1COL]{WileyNJDv5} 
\articletype{Article Type}%

\received{Date Month Year}
\revised{Date Month Year}
\accepted{Date Month Year}
\journal{Journal}
\volume{00}
\copyyear{2023}
\startpage{1}

\usepackage{amsmath}
\usepackage{amsthm}
\usepackage{amssymb}
\usepackage{natbib}

\DeclareMathOperator*{\argmin}{arg\,min}

\begin{document}

\title{A corrected score function framework for modelling circadian gene expression}

\author[1]{Michael T. Gorczyca}

\author[2]{Tavish M. McDonald}

\author[3]{Justice D. Sefas}

\authormark{GORCZYCA \textsc{et al.}}

\address[1]{\orgname{MTG Research Consulting}, \orgaddress{\city{Pittsburgh}, \state{Pennsylvania}, \country{United States of America}}}

\address[2]{\orgdiv{Computational Engineering Division}, \orgname{Lawrence Livermore National Laboratory}, \orgaddress{\city{Livermore}, \state{California}, \country{United States of America}}}

\address[3]{\orgdiv{Department of Computer Science}, \orgname{University of British Columbia}, \orgaddress{\city{Vancouver}, \state{British Columbia}, \country{Canada}}}

\corres{Corresponding author Michael T. Gorczyca. \email{mtg62@cornell.edu}}

\abstract[Abstract]{Many biological processes display oscillatory behavior based on an approximately 24 hour internal timing system specific to each individual. One process of particular interest is gene expression, for which several circadian transcriptomic studies have identified associations between gene expression during a 24 hour period and an individual's health. A challenge with analyzing data from these studies is that each individual's internal timing system is offset relative to the 24 hour day-night cycle, where day-night cycle time is recorded for each collected sample. Laboratory procedures can accurately determine each individual's offset and determine the internal time of sample collection. However, these laboratory procedures are labor-intensive and expensive. In this paper, we propose a corrected score function framework to obtain a regression model of gene expression given internal time when the offset of each individual is too burdensome to determine. A feature of this framework is that it does not require the probability distribution generating offsets to be symmetric with a mean of zero. Simulation studies validate the use of this corrected score function framework for cosinor regression, which is prevalent in circadian transcriptomic studies. Illustrations with three real circadian transcriptomic data sets further demonstrate that the proposed framework consistently mitigates bias relative to using a score function that does not account for this offset.}


\maketitle

\renewcommand\thefootnote{\fnsymbol{footnote}}
\setcounter{footnote}{1}

\section{Introduction}

The circadian clock is an internal timekeeping mechanism present in organisms across disparate taxa, and is responsible for regulating biological processes ranging from gene expression to organism behavior \citep{Roenneberg2016}. In humans, these clocks typically operate relative to the 24 hour day-night cycle (Zeitgeber time, or ZT) through an entrainment process, where external environmental cues associated with the day-night cycle enforce synchrony between an individual's circadian clock time (internal circadian time, or ICT) and ZT \citep{Lewy1999, Ruiz2020, Wittenbrink2018, Wright2013}. However, each individual's ICT can be offset relative to ZT due to individual-specific factors, such as their genetic makeup \citep{Hsu2015}, their age \citep{Kennaway2023}, and their exposure to environmental conditions \citep{Allebrandt2014, Stothard2017, Wright2013}. To be precise, each individual's circadian clock operates on an approximately 24 hour period, but the ICT of each individual is offset when compared to ZT \citep{Duffy2011, Wittenbrink2018}. 

The offset of ICT relative to ZT presents a challenge in analyzing circadian transcriptomic study data, as the recorded time of observed gene expression is ZT \citep{Ceglia2018}. Laboratory procedures can be performed on each individual to determine their ZT when melatonin onset occurs in dim-light conditions, or DLMO time, which is a gold standard marker for this offset \citep{Hughey2017, Lewy1999, Ruiz2020, Wittenbrink2018}. However, many procedures for accurately determining DLMO time are labor-intensive and expensive, as they involve taking multiple samples from an individual over several hours under controlled conditions \citep{Kantermann2015, Reid2019}.

This paper is motivated by the challenge of determining ICT, and considers the offset of ICT relative to ZT akin to covariate measurement error. Several frameworks in statistics \citep{Buonaccorsi2010, Carroll2006, Fuller1987} and in econometrics \citep{Chen2011, Schennach2012} have been proposed to account for covariate measurement error in regression modelling. We propose a corrected score function framework that uses the moments of the probability distribution generating measurement error to obtain consistent estimators and unbiased inferences. Notably, several corrected score function frameworks assume covariate measurement error is generated by a symmetric probability distribution with a mean of zero \citep{Cai2021, Hong2003, Nakamura1990, Nakamura1992, Stefanski1989}. In this paper, the framework proposed is able to accommodate asymmetric distributions with a non-zero mean.

The remainder of this paper is organized as follows. In Section \ref{sec:2}, the corrected score function framework for estimation and inference is presented. In Section \ref{sec:3}, a simulation study is performed to compare this corrected score function to a score function that does not account for measurement error. In Section \ref{sec:4}, illustrations with this corrected score function framework are presented with three real circadian transcriptomic data sets. Finally, in Section \ref{sec:5}, the results from this study and directions for future work are discussed.

\section{Methodology} \label{sec:2}

\subsection{Corrected score function overview} \label{sec:2.1}

Suppose $n$ independent realizations, $\{(X^{\dagger}_j, Y_j), \  j=1,\ldots,n\}$, of the random variables $(X^{\dagger}, Y)$ are collected during a circadian biology experiment. Here, $X^{\dagger} = X^{*} +\xi$ represents the mismeasured covariate (Zeitgeber time, or ZT, multiplied by $\pi/12$), and $\xi$ represents the measurement error (ZT of melatonin onset under dim-light conditions, or DLMO time, multiplied by $\pi/12$) of the true covariate $X^{*}$ (internal circadian time, or ICT, multiplied by $\pi/12$). The measurement error $\xi$ is assumed to be independent of both $X^{*}$ and the response $Y$ (gene expression), which is described in literature as a classical measurement error problem \citep{Carroll2006}.

In many circadian biology studies \citep{Arnardottir2014, Cornelissen2014, delolmo2022, Li2013, Lundell2020}, the cosinor regression model is assumed to be correctly specified for the response $Y$, or
\begin{align} \label{eq:1}
    Y =& (\theta^{*})^Tf(X^{*}) + \epsilon =\theta^{*}_0 + \theta^{*}_{1}\text{\text{sin}}( X^{*}) + \theta^{*}_{2}\text{\text{\text{cos}}}(X^{*}) + \epsilon, 
\end{align}
where $\theta^{*}$ denotes a vector of unknown model parameters, and
\begin{align*}
    f(Z) =&  \left\{1, \text{sin}(Z), \text{cos}(Z)\right\}^T
\end{align*}
is a vector of regression functions with a univariate argument $Z$. The random noise $\epsilon$ is independent of $X^{\dagger}$, $X^*$, and $\xi$, and is generated from a probability distribution with a mean of zero and a finite, constant variance. An alternate parameterization of the cosinor regression model from (\ref{eq:1}) commonly analyzed in circadian biology studies is its amplitude-phase representation, or
\begin{equation} \label{eq:3}
    Y = \theta^{*}_0 + A^*\text{\text{cos}}( X^{*}+\eta^{*}) + \epsilon.
\end{equation}
Here, $A^*$ is biologically interpreted as the extent at which gene expression levels vary from their mean expression levels, and $\eta^*$ as the time at which gene expression levels peak. The relationships between parameters in (\ref{eq:1}) and (\ref{eq:3}) are given by
\begin{equation}
   \theta^*_{1} = -A^*\sin(\eta^*), \hspace{0.35em} \theta^*_{2} = A^*\cos(\eta^*), \hspace{0.35em} A^* = \sqrt{(\theta^*_{1})^2+(\theta^*_{2})^2}, \hspace{0.35em} \eta^* = \text{atan2}(-\theta^*_{1}, \theta^*_{2}), \label{eq:alt_to_orig}
\end{equation}
where $\text{atan2}(Z_1, Z_2)$ denotes the two-argument arctangent function \citep{Cornelissen2014}.

In this paper, it is assumed that parameter vector estimation for the cosinor regression model is performed by minimizing squared loss \citep{Cornelissen2014, Zong2023}, or 
\begin{align*}
\hat{\theta} =& \argmin_{\theta \in \Theta} \ \ 
\frac{1}{n}\sum_{j=1}^n\mathcal{L}(Y_j, X_j^{\dagger}; \theta) = \argmin_{\theta \in \Theta} \ \ \frac{1}{n}\sum_{j=1}^n\left\{Y_j-\theta^Tf(X_j^{\dagger})\right\}^2,
\end{align*}
where $\hat{\theta}$ denotes the parameter vector estimate that minimizes squared loss, and $\Theta\subseteq \mathbb{R}^3$ denotes a compact subset of a Euclidean space. The gradient of squared loss with respect to $\theta$, or score function if $\epsilon$ is generated by a normal distribution, is defined as
\begin{align*}
\frac{1}{n}\sum_{i=1}^n g(Y_j, X^{\dagger}_j; \theta)=&\frac{1}{n}\sum_{j=1}^n\nabla \mathcal{L}(Y_j, X^{\dagger}_j; \theta) =-\frac{2}{n}\sum_{j=1}^nf(X^{\dagger}_j)Y_j - f(X^{\dagger}_j)f(X^{\dagger}_j)^T\theta, 
\end{align*}
where the parameter vector estimate $\hat{\theta}$ satisfies the condition 
\begin{align} \label{eq:emp_grad}
\frac{1}{n}\sum_{j=1}^ng(Y_j, X^{\dagger}_j; \hat{\theta})=0.
\end{align}

When the observed covariates are mismeasured, estimators obtained from a score function that does not account for measurement error are generally inconsistent. In the context of classical measurement error, this mismeasurement results in attenuation bias, where the estimated parameters are biased toward the direction of zero relative to the true parameters \citep{Carroll2006, Griliches1970, Meijer2021}. The corrected score function framework is one approach to mitigating this attenuation bias, where the objective is to identify a score function $\tilde{g}(Y, X^{\dagger}; \theta)$ such that $\mathbb{E}\{g(Y, X^{*}; \theta)\}=\mathbb{E}\{\tilde{g}(Y, X^{\dagger}; \theta)\}$ for each $\theta \in \Theta$. The following theorem presents a general corrected score function for regression given a single mismeasured covariate. 
\newline
\begin{theorem} \label{thm1} Suppose $X^{\dagger} = X^{*}+\xi$, where $X^*$ and $\xi$ are independent. Assume:
\begin{enumerate}
    \item The characteristic function for $\xi$ is smooth and is non-zero for all arguments. 
    \item For each fixed $Z_1, Z_2 \in \mathbb{R}$, $g(Z_1, Z_2; \theta)$ is continuous in $\theta$. Further,
    \begin{align*}
\mathbb{E}\{\sup_{\theta \in\Theta}||g(Y, X^{\dagger}; \theta)||_2\} = \mathbb{E}\{\sup_{\theta \in\Theta}||g(Y, X^*+\xi; \theta)||_2\} <\infty, 
\end{align*}
and the summation $$\sum_{m=0}^{\infty}\left|\left|\frac{\kappa^{(m)}(0)}{m!}\mathbb{E}\left\{\frac{d^mg(Y, X^{\dagger}; \theta)}{d(X^{\dagger})^m}\right\}\right|\right|_2 < \infty,$$
where $\kappa(t)=1/\mathbb{E}\{\exp(t\xi)\}$, and $\kappa^{(m)}(0)$ denotes the $m$-th order derivative of $\kappa(t)$ evaluated at $t=0$.
\end{enumerate}
Then for each $\theta \in \Theta$,
\begin{align*}
    \mathbb{E}\{g(Y, X^{*}; \theta)\}=\mathbb{E}\{\tilde{g}(Y, X^{\dagger}; \theta)\} =\sum_{m=0}^{\infty}\frac{\kappa^{(m)}(0)}{m!}\mathbb{E}\left\{\frac{d^mg(Y, X^{\dagger}; \theta)}{d(X^{\dagger})^m} \right\}. 
\end{align*}
\end{theorem}

The derivation for this result is provided in Appendix \ref{app:A}. The first assumption in Theorem \ref{thm1} is common in measurement error literature \citep{Belomestny2021}, and ensures that each moment of the probability distribution for measurement error is finite. The second assumption guarantees that the corrected score function derived satisfies absolute convergence \citep{Hong2003, Stefanski1989}, which ensures consistent estimators can be obtained. 

There are two implications of Theorem \ref{thm1}. The first implication is one can use this corrected score function to identify a consistent estimator based on whether or not it satisfies the equality
\begin{align}\label{eq:cor_cond}
\frac{1}{n}\sum_{i=1}^n\sum_{m=0}^{\infty}\frac{\kappa^{(m)}(0)}{m!}\left\{\frac{d^mg(Y_i, X^{\dagger}_i; \hat{\theta})}{d(X^{\dagger})^m}\right\} =0
\end{align}
when the covariate data are mismeasured and estimates of each $\kappa^{(m)}(0)$ are known. The second implication is that the corrected score function makes general assumptions about the probability distribution generating measurement error. In particular, several research efforts assume specific probability distributions that are symmetric with a mean of zero \citep{Cai2021, Hong2003, Nakamura1990, Nakamura1992, Stefanski1989}. The corrected score function in (\ref{eq:cor_cond}) accommodates an asymmetric probability distribution with a non-zero mean when warranted. 

\subsection{Inference in the presence of measurement error} \label{sec:2.4}

One consequence of using mismeasured covariate data for estimation is that it cascades to a loss of statistical power for hypothesis testing \citep{Devine1998, Lagakos1988, Mckeown1994, Tosteson1988, White1994}. For example, a hypothesis test widely used in circadian transcriptomic studies is the $F$-test \citep{Cornelissen2014, Zong2023}, where an investigator interprets the $p$-value computed from this test as the significance of a gene's oscillatory behavior over a 24 hour period \citep{Ding2021}. An asymptotic version of the $F$-test for illustration is a Wald test with the null hypothesis $H_0:\hat{\theta} = \hat{\theta}^{\text{(Null)}}$, where 
\begin{align*}
\hat{\theta}^{\text{(Null)}} = \left(\frac{1}{n}\sum_{j=1}^nY_j,0,0 \right)
\end{align*}
represents the empirical parameter vector obtained when an intercept model is specified for estimation \citep{Davidson2003, Meijer2021}. The Wald test statistic computed for this hypothesis is
\begin{align*}
\tau_{\text{Wald}} = \frac{n}{\hat{\sigma}^2}(\hat{\theta}-\hat{\theta}^{\text{(Null)}})^T\left\{\frac{1}{n}\sum_{j=1}^nf(X^{\dagger}_j)f(X^{\dagger}_j)^T\right\}(\hat{\theta}-\hat{\theta}^\text{(Null)}),
\end{align*}
where $\hat{\sigma}^2$ is the empirical estimate of the squared residual error based on the estimated parameter vector $\hat{\theta}$. The test statistic $\tau_{\text{Wald}}$ computed from a cosinor regression model from (\ref{eq:1}) is assumed to follow a central chi-squared distribution with $2$ degrees of freedom, and the null hypothesis is rejected at a pre-determined $\alpha$-level if $\tau_F$ surpasses the $1-\alpha$ percentile of the central chi-squared distribution \citep{Boos2013}.

When performing this Wald test with mismeasured covariate data, a loss of statistical power can stem from two components. The first component is the estimated parameter vector $\hat{\theta}$, which suffers from attenuation bias. The second component is the squared residual error, which exhibits inflation bias, or is larger than the squared residual error computed from correctly measured covariate data \citep{Carroll2006, Meijer2021}. To address this loss of statistical power, measurement error research efforts rely on score test statistics for hypothesis testing \citep{Hong2003, Stefanski1991, Tosteson1988, Tosteson2003}, which enables use of the corrected score function from Theorem \ref{thm1}. The assumption of correct model specification in (\ref{eq:1}) implies that the score test statistic for the hypothesis test $H_0:\hat{\theta} = \hat{\theta}^{\text{(Null)}}$ is
\begin{align} 
    \tau_{\text{score}} =& n\left\{\frac{1}{n}\sum_{j=1}^n\tilde{g}(Y_j, X^{\dagger}_j; \hat{\theta}^{(\text{Null})})\right\}^T\left\{\frac{1}{n}\sum_{j=1}^n\frac{d\tilde{g}(Y_j, X^{\dagger}_j; \hat{\theta}^{(\text{Null})})}{d\theta}\right\}^{-1} \nonumber \\
    & \quad \times \left\{\frac{1}{n}\sum_{j=1}^n\tilde{g}(Y_j, X^{\dagger}_j; \hat{\theta}^{(\text{Null})})\right\}, \label{eq:st_t}
\end{align} 
where $d\tilde{g}(Y, X^{\dagger}; \theta)/d\theta$ is the derivative of the corrected score function for the cosinor regression model with respect to $\theta$ \citep{Boos2013}. The asymptotic distribution of the score test statistic in (\ref{eq:st_t}) under the null hypothesis is also a central chi-squared distribution with $2$ degrees of freedom for a cosinor regression model. This null hypothesis is rejected at a pre-determined $\alpha$-level if $\tau_{\text{Score}}$ surpasses the $1-\alpha$ percentile of the central chi-squared distribution \citep{Boos2013, Tosteson2003}. 

\section{Simulation Study}  \label{sec:3}

A simulation study was conducted to evaluate the utility of the corrected score function from Theorem \ref{thm1} for estimating a cosinor regression model from (\ref{eq:1}) and performing inference with an estimated model. This study considers eight simulation scenarios for generating the true parameters for the amplitude-phase representation of a cosinor regression model in (\ref{eq:3}), measurement error, observed covariate, and random noise. It is emphasized that the first four scenarios will generate data under the assumptions of a classical measurement error problem, whereas the last four scenarios will generate data under the assumptions of a Berkson measurement error problem where $X^* = X^{\dagger}+\xi$ with $X^{\dagger}$ and $\xi$ independent \citep{Berkson1950}. The motivation for these last four scenarios with Berkson measurement error is to assess the utility of the corrected score function from Theorem \ref{thm1} when the assumption of classical measurement error is violated. To accommodate these two types of measurement errors in describing the simulation scenarios, $X_j$ will denote the generated covariate that is independent of the measurement error $\xi_j$. Specifically, for the first four simulation scenarios, $X_j$ represents $X^*_j$ (a classical measurement error problem). For the last four simulation scenarios, $X_j$ represents $X_j^{\dagger}$ (a Berkson measurement error problem). The response variable is then generated following (\ref{eq:1}). 
\begin{description}
    \item[Scenarios 1 and 5.] $\theta^*_0=6$, $A^*= 1$, $\eta^* = 0$, $\xi_j \sim \text{wL}(0, \sqrt{0.3})$, $X_j=2\pi(j-1)/n$, and $\epsilon_j \sim \text{N}(0, 0.25)$.  
    \item[Scenarios 2 and 6.] $\theta^*_0 \sim \text{TN}(6, 1, 4, 8)$, $A^*= 0.5$, $\eta^* = \pi/4$, $\xi_j \sim \text{wSN}(0.5, 0.5, -0.7)$, $X_j=2\pi(j-1)/n$, and $\epsilon_j \sim \text{N}(0, 1)$.
    \item[Scenarios 3 and 7.] $\theta^*_0 = 6$, $A^* \sim \text{TN}(0.5, 0.2, 0.2, 0.8)$, $\eta^*=\pi/2$, $\xi_j \sim \text{wL}(0, \sqrt{0.3})$, $X_j \sim \text{wN}(\pi, 2.5)$, and $\epsilon_j \sim \text{N}(0, 0.25)$. \label{set:4} 
    \item[Scenarios 4 and 8.] $\theta^*_0\sim \text{TN}(6, 1, 4, 8)$, $A^* \sim \text{TN}(1, 0.2, 0.7, 1.3)$, $\eta^*=5\pi/4$, $\xi_j \sim \text{wSN}(-0.5, 0.5, 0.7)$, $X_j\sim \text{wN}(\pi, 2.5)$, and $\epsilon_j \sim \text{N}(0, 1)$. \label{set:5} 
 \end{description}
For each simulation scenario, $\text{N}(\mu, \sigma^2)$ denotes a normal distribution with mean $\mu$ and variance $\sigma^2$; $\text{TN}(\mu, \sigma^2, a, b)$ a truncated normal distribution with mean $\mu$, variance $\sigma^2$, lower bound $a$, and upper bound $b$; $\text{wL}(\mu, b)$ a wrapped Laplace distribution with location $\mu$ and scale $b$; $\text{wN}(\mu, \sigma^2)$ a wrapped normal distribution with mean $\mu$ and variance $\sigma^2$; $\text{wSN}(\mu, \sigma^2, \gamma)$ a wrapped skew normal distribution with mean $\mu$, variance $\sigma^2$, and skewness $\gamma$. 

Scenarios 1, 3, 5, and 7 represent a situation where the covariate data are generated from an equispaced experimental design, or when $X_j = 2\pi(j-1)/n$. Equispaced designs are optimal for obtaining information from a cosinor regression model under multiple statistical criteria \citep{Federov1972, Pukelsheim2006} and are advocated in circadian transcriptomic studies \citep{Hughes2017, Zong2023}. Scenarios 2, 4, 6, and 8, on the other hand, represent a situation where the data are collected from an observational study. Additionally, Scenarios 1, 4, 5, and 8 represent a situation where the amplitude is relatively weak, and Scenarios 2, 3, 6, and 7 represent a situation where the amplitude is relatively strong. It is emphasized that Scenarios 2-4 and 6-8 also represent settings where there are individual-specific variations in gene expression \citep{Cheung2003, Rockman2006, Stranger2007}. To be precise, in Scenarios 2 and 6 the mean expression level for a gene is different for each individual, but the variation in amplitude is the same. In Scenarios 3 and 7 each individual has the same mean expression level for a gene, but the variation in amplitude is different. Finally, in Scenarios 4 and 8 the mean expression level for a gene and the variation in amplitude are different for each individual.

For each simulation scenario, 2,000 simulation trials are performed. In each simulation trial, three data sets are generated. These data sets consist of $n=100$, $n=400$, and $n=1,600$ samples, which represent settings where a small \citep{Braun2018}, moderate \citep{Arnardottir2014}, and large \citep{gtex2020} number of samples are available, respectively. In each simulation trial, two estimation frameworks are considered for cosinor regression:
\begin{description}
    \item[Framework 1.] A parameter vector estimate that satisfies (\ref{eq:cor_cond}) is obtained from mismeasured covariate data.
    \item[Framework 2.] A parameter vector estimate that satisfies (\ref{eq:emp_grad}) is obtained from mismeasured covariate data.
 \end{description}
It is emphasized that the corrected score function in Framework 1 is an infinite summation. To address this, estimation and inference with Framework 1 will use the first three terms ($m=0, 1, 2$) of the corrected score function. Additionally, each $\kappa^{(m)}(0)$ required for the corrected score function in Framework 1 is computed from the generated $\xi_j$ in each simulation trial for Scenarios 1-4, and $-\xi_j$ for Scenarios 5-8. The use of $-\xi_j$ in Scenarios 5-8 accounts for the fact that $X^{\dagger}=X^{*}-\xi$ in a Berkson measurement error problem, whereas $X^{\dagger} = X^{*}+\xi$ in a classical measurement error problem. It is noted that the sample sizes imply differences in knowledge about each $\kappa^{(m)}(0)$: $n=100$ represents a situation where each $\kappa^{(m)}(0)$ is known with relatively low precision, $n=400$ indicates moderate precision, and $n=1,600$ signifies high precision.

After conducting 2,000 simulation trials, the median and standard deviation of the following quantities as well as their absolute values are reported:
\begin{itemize}
    \item[1.] $(\hat{\theta}_0-\theta^{*}_0)$ or the difference between the estimated intercept parameter for the model defined in (\ref{eq:3}) and the true intercept parameter.
    \item[2.] $(\hat{A}-A^{*})/A^{*}$, or the difference between the estimated amplitude from the amplitude-phase representation defined in (\ref{eq:3}) and the true amplitude divided by the true amplitude.
    \item[3.] $\hat{\eta}-\eta^{*}$, or the difference between the estimated phase-shift from the amplitude-phase representation defined in (\ref{eq:3}) and the true phase-shift.
    \item[4.] $(\tau_{\text{Score}}-\tau^*_{\text{Score}})/n$, or the difference between the estimated score test statistic from (\ref{eq:st_t}) computed with the estimated parameters and the corresponding quantity computed with the true parameters scaled by sample size.
\end{itemize} 
The calculation of $\tau_{\text{Score}}$ in Framework 2 replaces $\tilde{g}(Y, X^{\dagger}; \theta)$ and its derivative $d\tilde{g}(Y, X^{\dagger}; \theta)/d\theta$ from (\ref{eq:st_t}) with $g(Y, X^{\dagger}; \theta)$ and $dg(Y, X^{\dagger}; \theta)/d\theta$, respectively. Similarly, $\tau^*_{\text{Score}}$ is computed with correctly measured covariate data, or with $g(Y, X^*; \theta)$ and $dg(Y, X^*; \theta)/d\theta$. It is noted that the amplitude quantity is divided by $A^*$ to account for differences between amplitude magnitudes for Scenarios 1,4,5, and 7 (which are relatively small) relative to Scenarios 2, 3, 6, and 7 (which are relatively large). Further, when the amplitude parameter is individual-specific in Scenarios 3, 4, 7, and 8, the mean parameter for the distribution that generates amplitudes serves as the true parameter (specifically $A^*=0.5$ for Scenarios 3 and 7, and $A^*=1$ for Scenarios 4 and 8).

Table \ref{tab:sim1} presents the results for all eight simulation scenarios. When the covariate data are mismeasured, Framework 1 consistently mitigates bias in amplitude estimation and score test statistic calculation, as well as phase-shift estimation for Scenarios 2, 4, 6, and 8 when the probability distribution generating measurement error is asymmetric. The similar performance of Frameworks 1 and 2 in phase-shift estimation for Scenarios 1, 3, 5, and 7 empirically indicates that phase-shift estimates are consistent when the probability distribution generating measurement error is symmetric. Further, both frameworks obtain similar performance in estimating $\theta^*_0$. When Framework 2 outperforms Framework 1 in estimating $\theta^*_0$, the difference between the median errors obtained from Frameworks 1 and 2 never exceeds $0.02$ in magnitude, and the difference between median absolute errors obtained from the two frameworks never exceeds $0.01$ in magnitude.

\begin{figure*}[!h]
\centering
\includegraphics[scale=0.14]{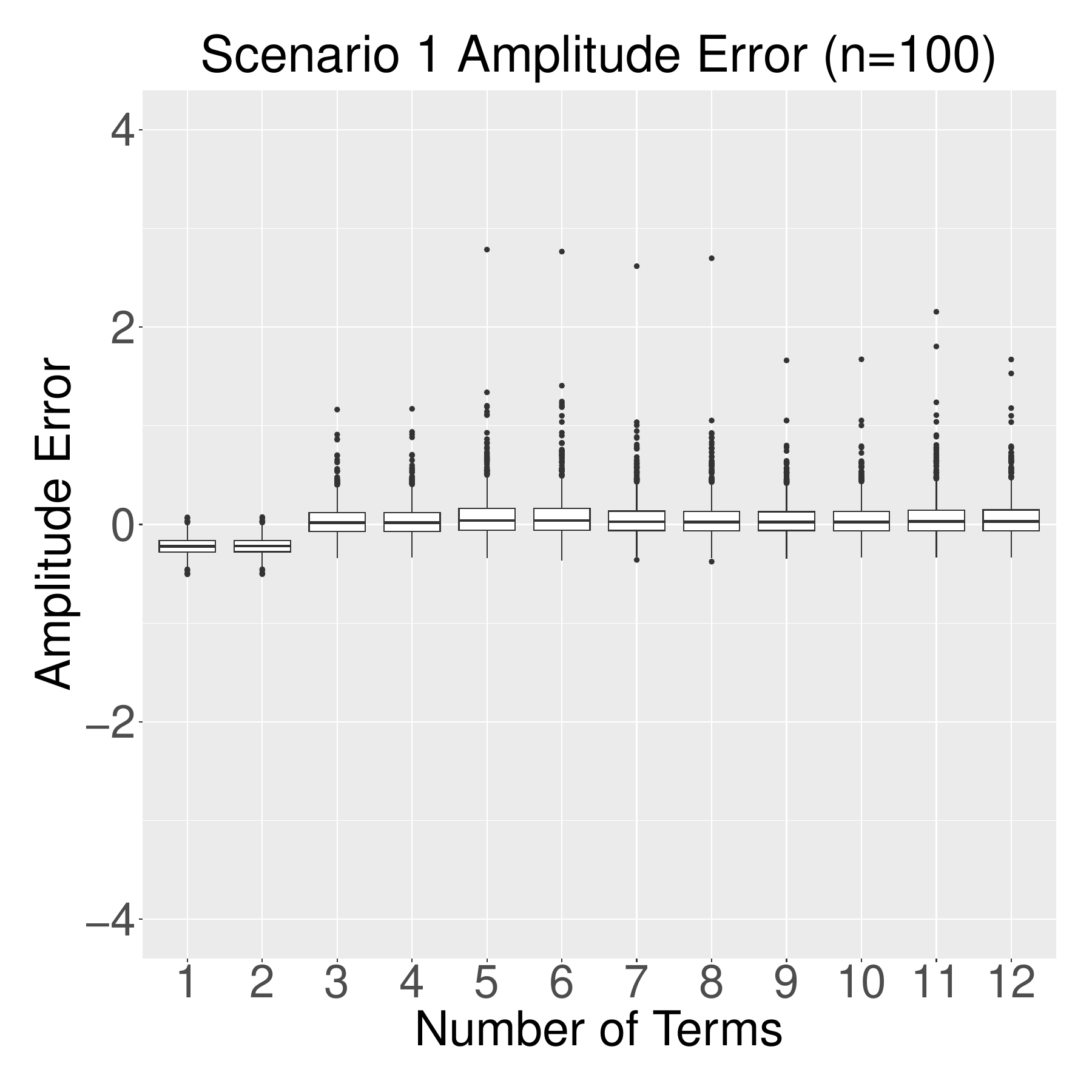}
\includegraphics[scale=0.14]{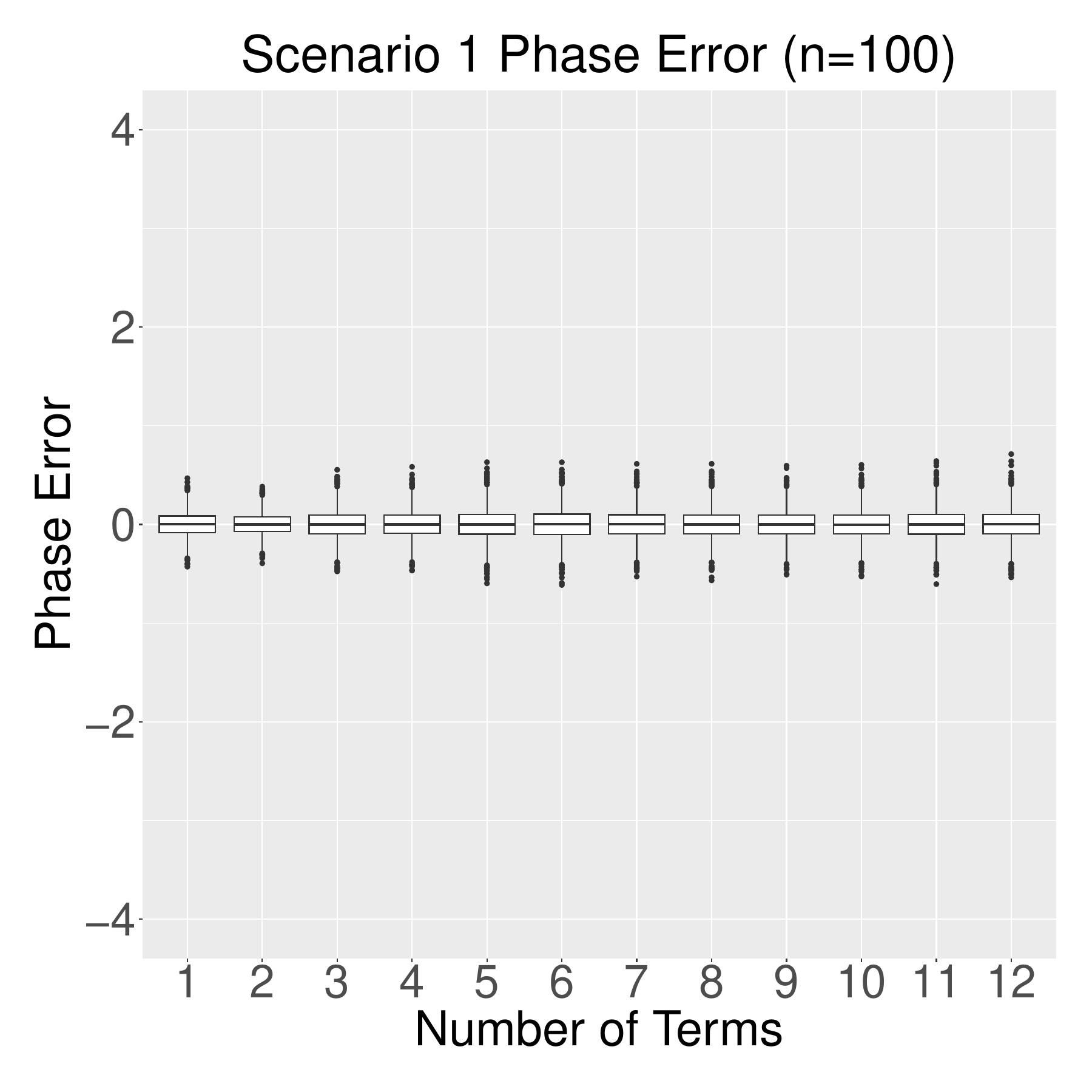}
\includegraphics[scale=0.14]{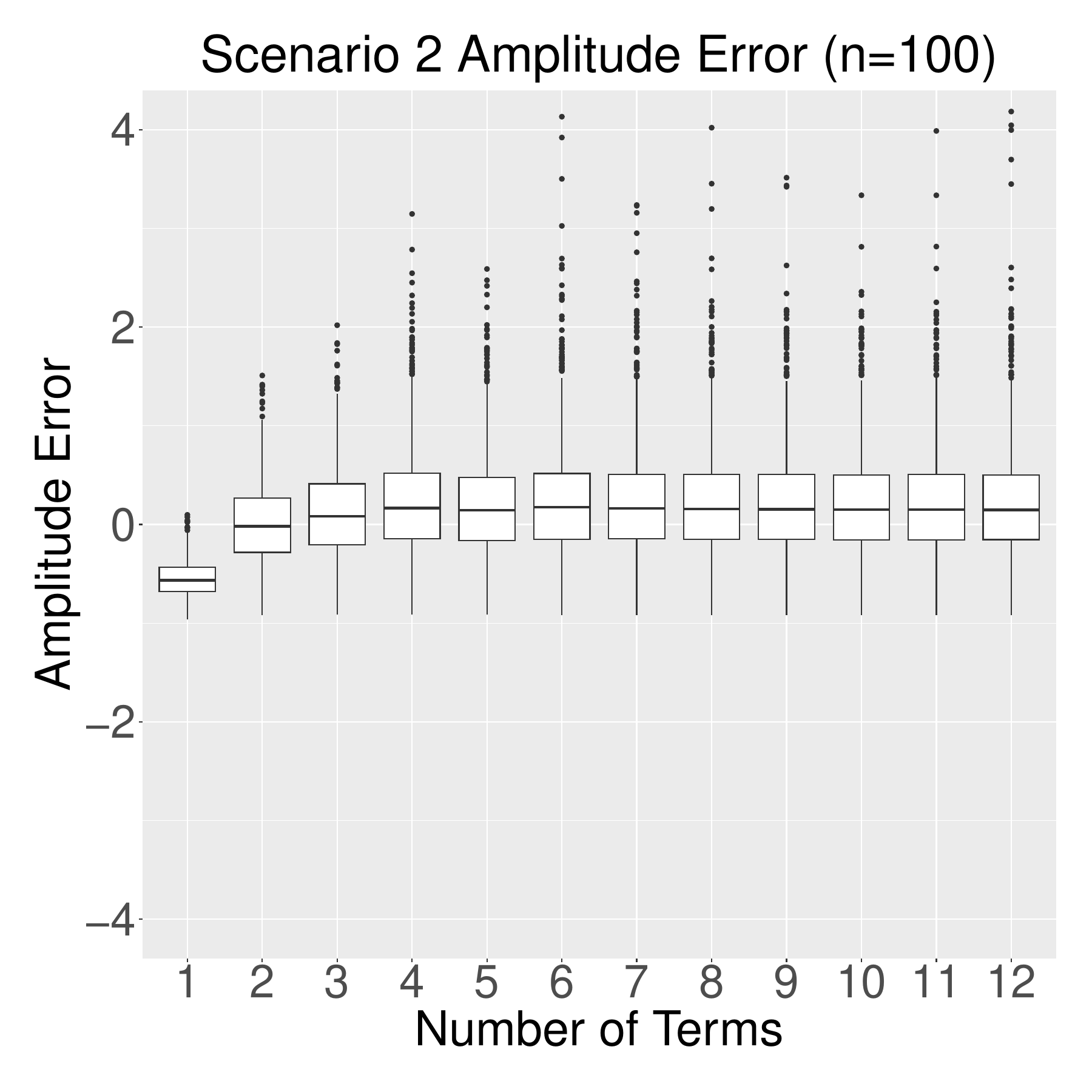}
\includegraphics[scale=0.14]{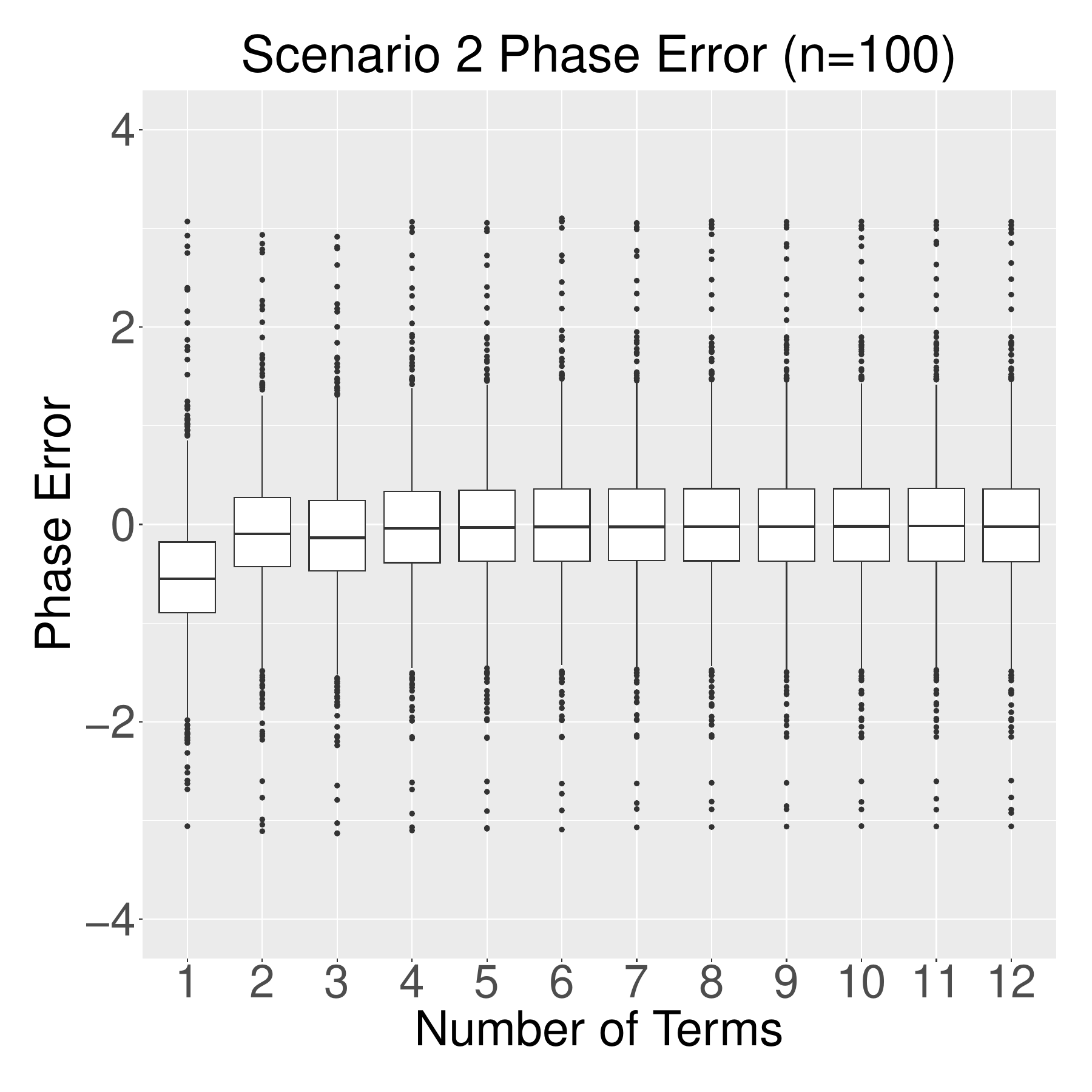}
\includegraphics[scale=0.14]{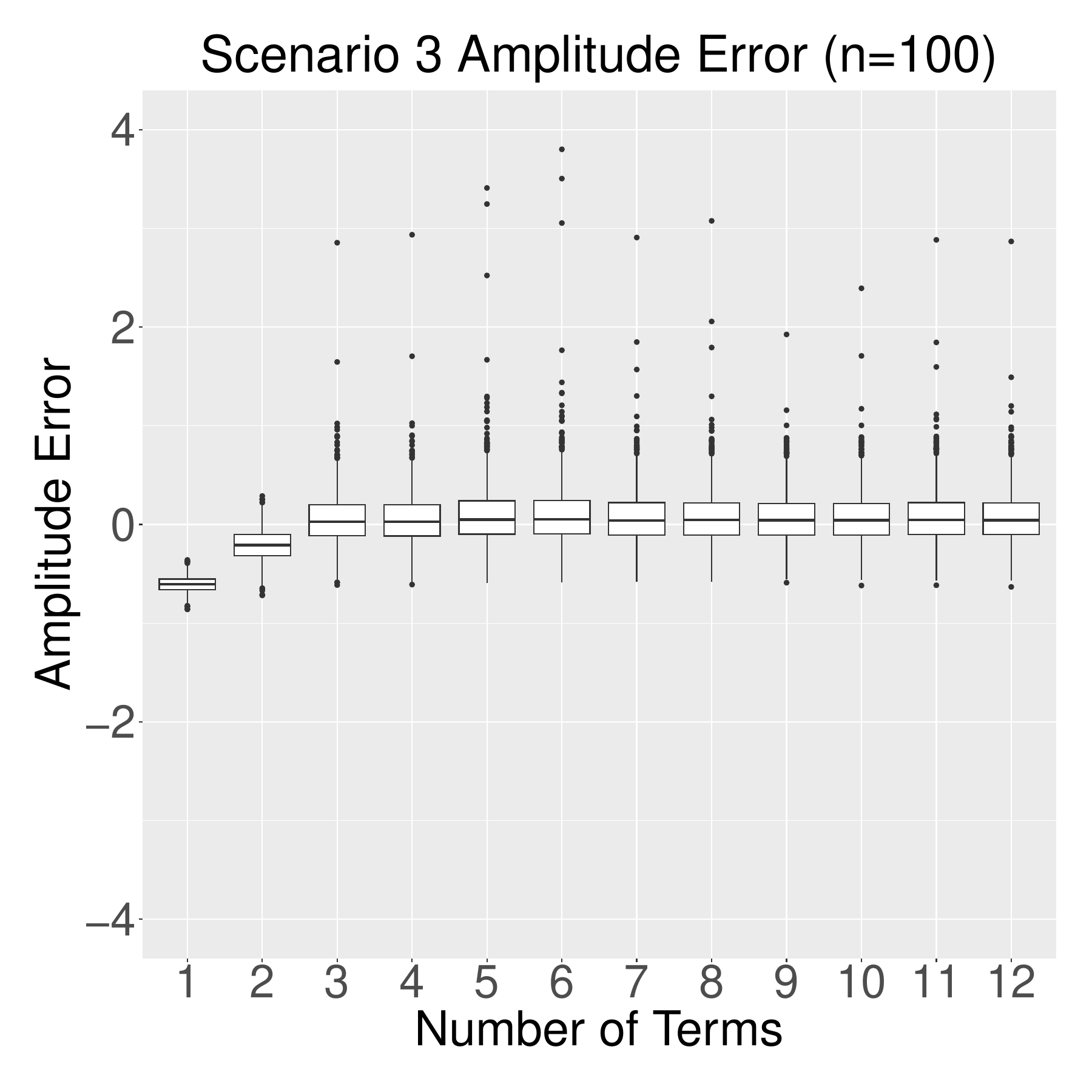}
\includegraphics[scale=0.14]{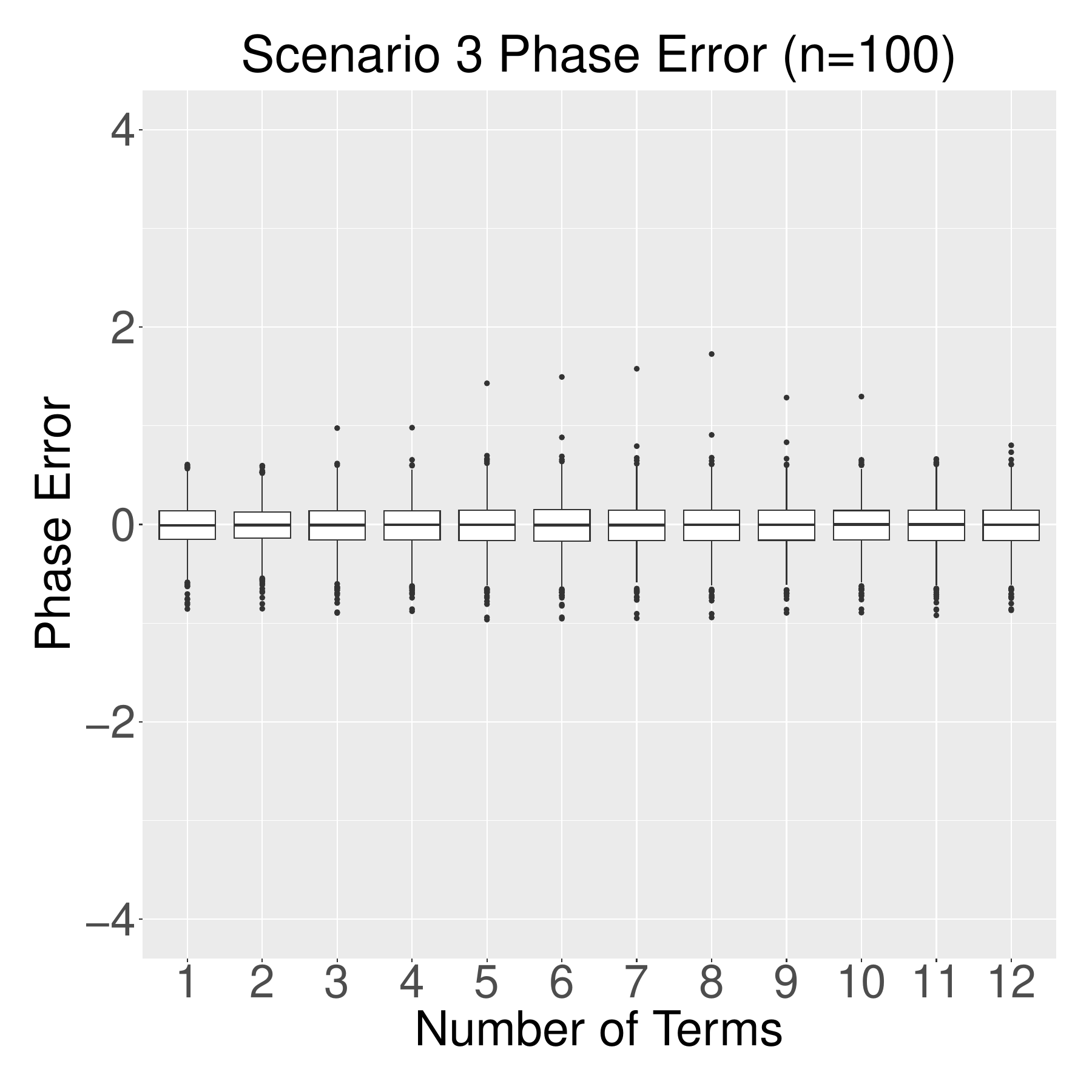}
\includegraphics[scale=0.14]{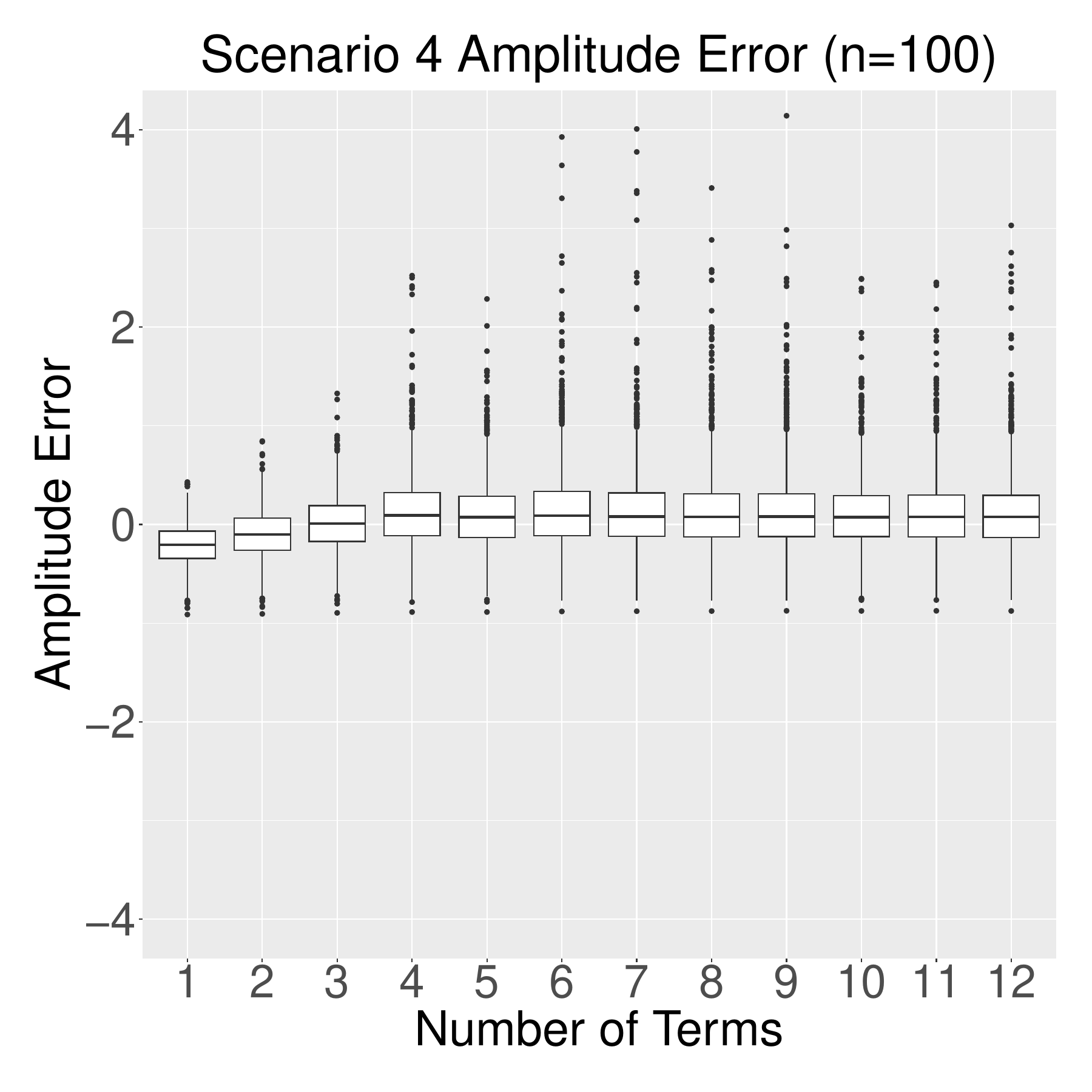}
\includegraphics[scale=0.14]{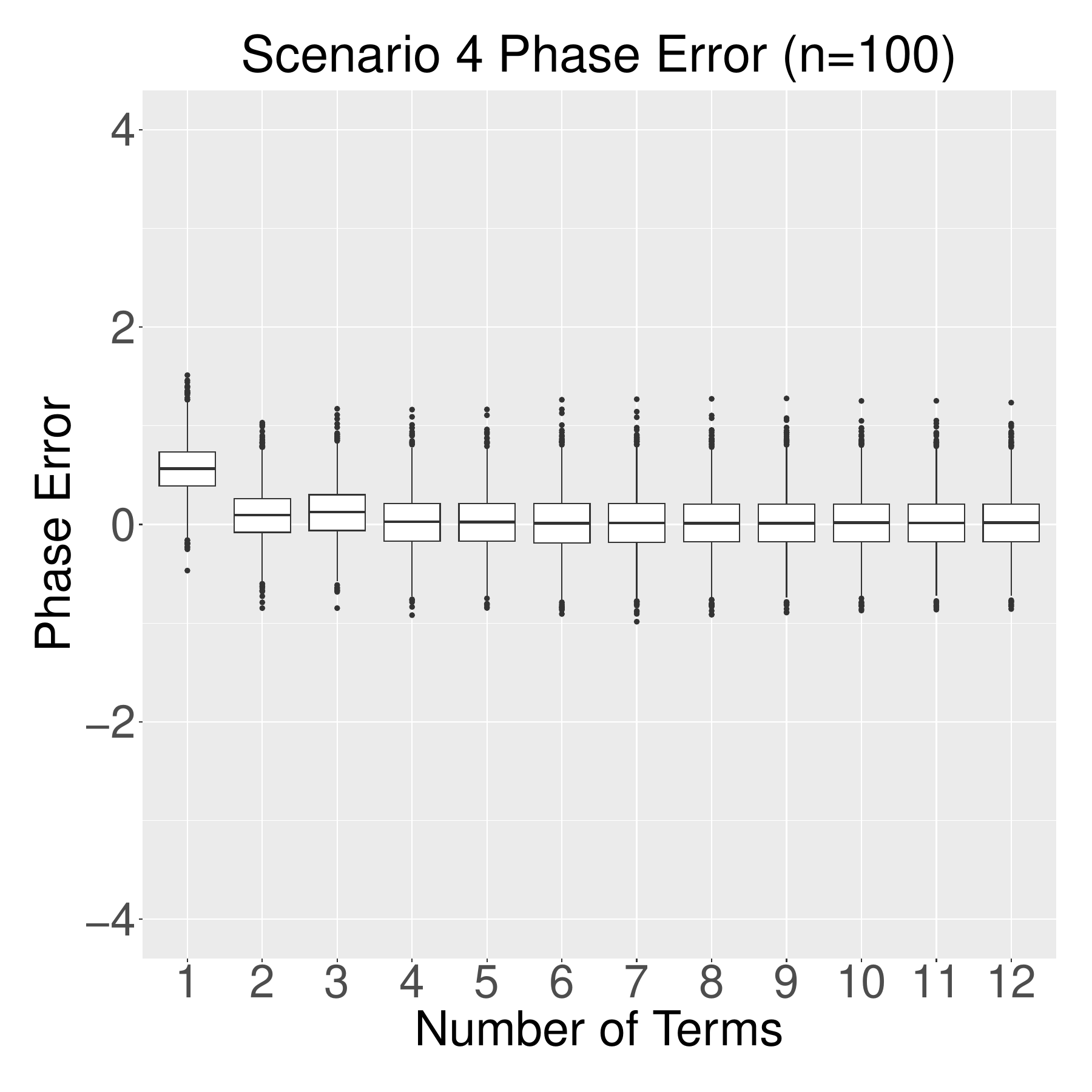}
\includegraphics[scale=0.14]{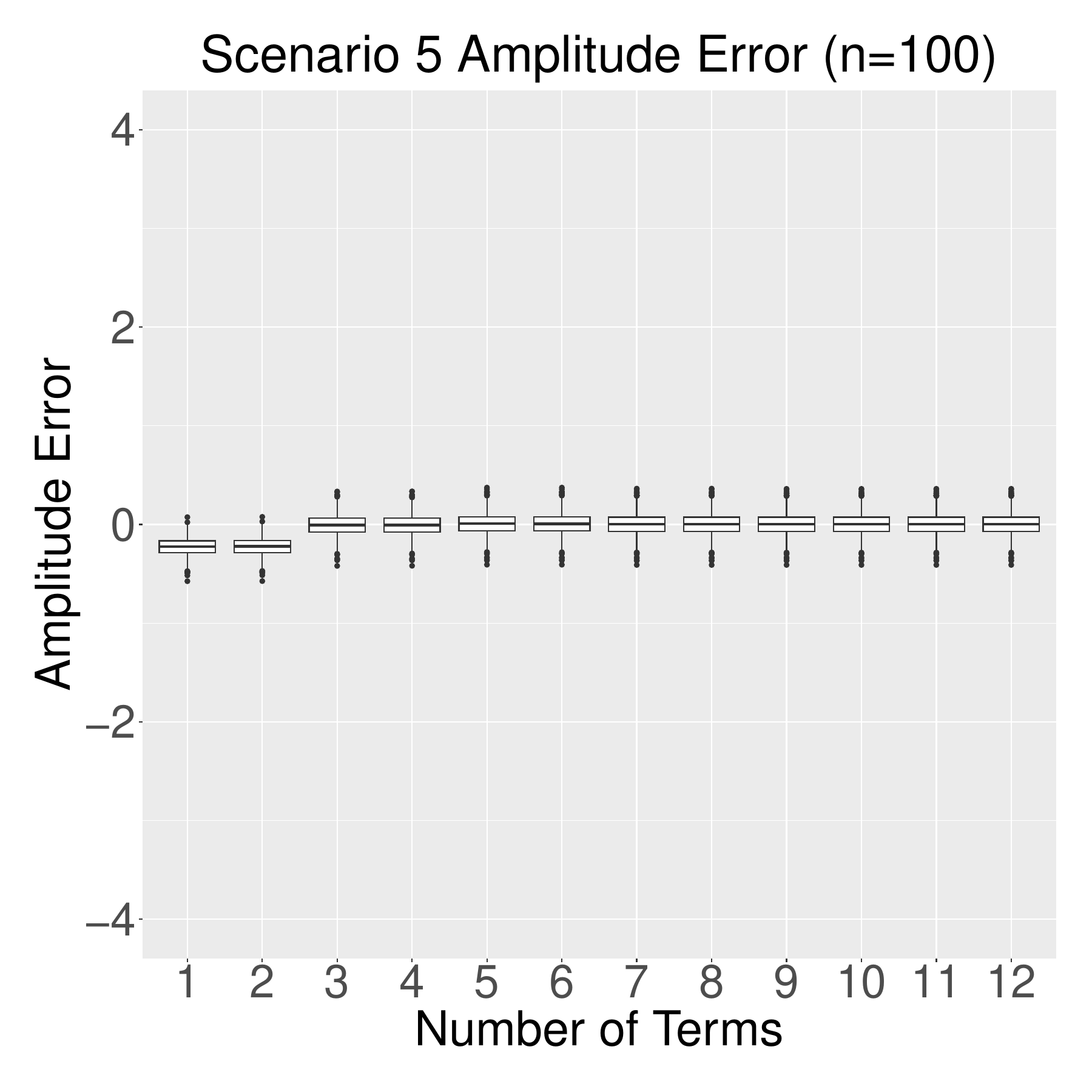}
\includegraphics[scale=0.14]{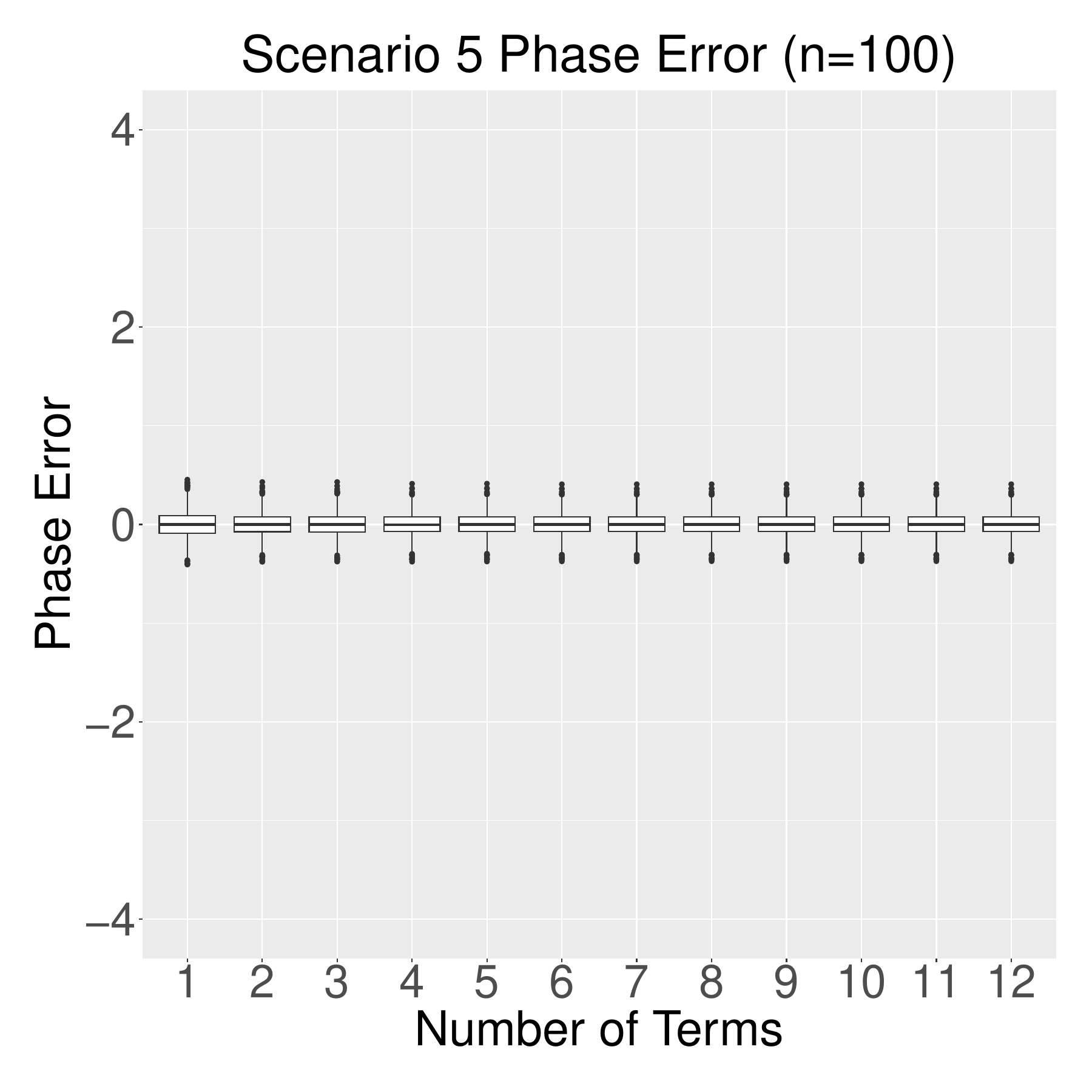}
\includegraphics[scale=0.14]{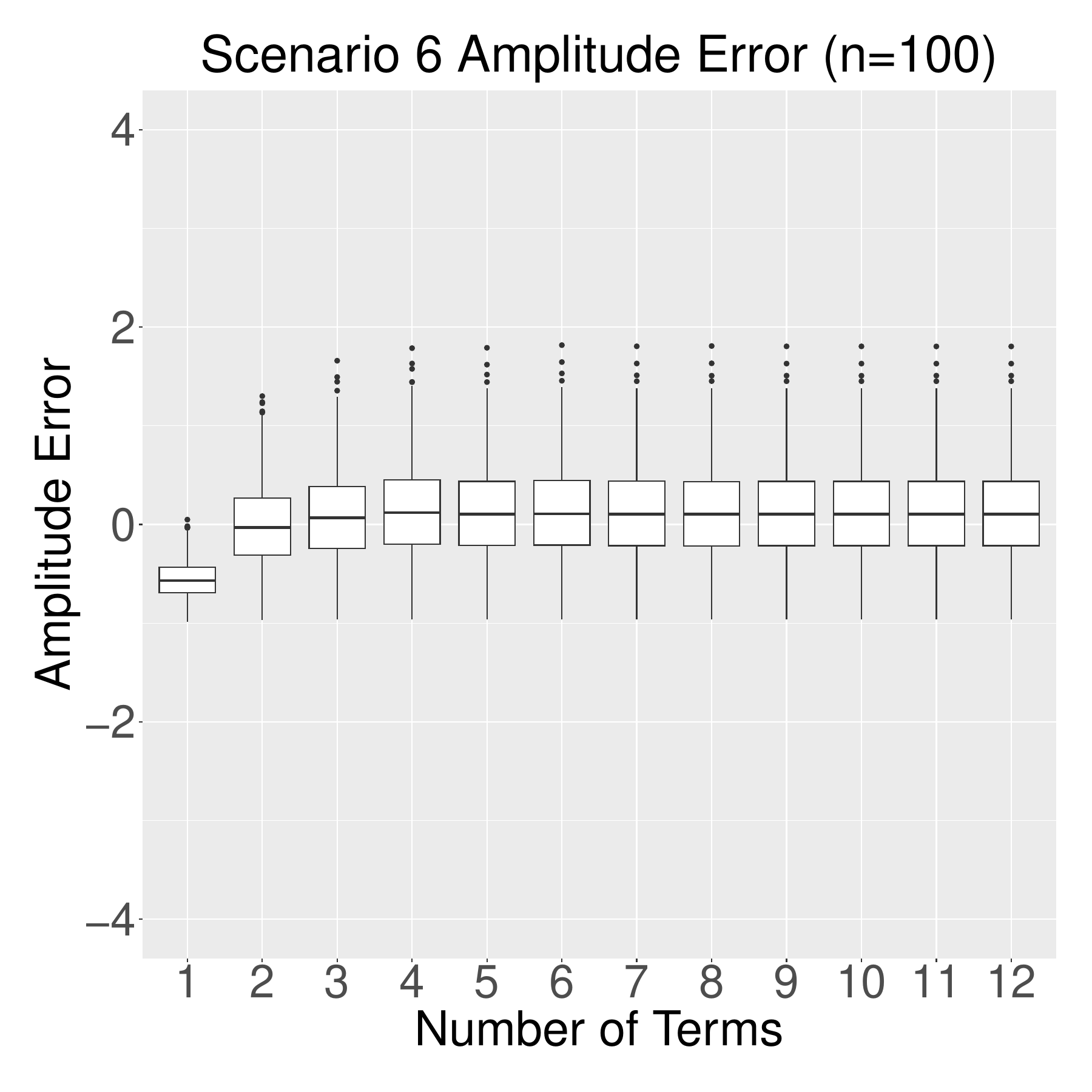}
\includegraphics[scale=0.14]{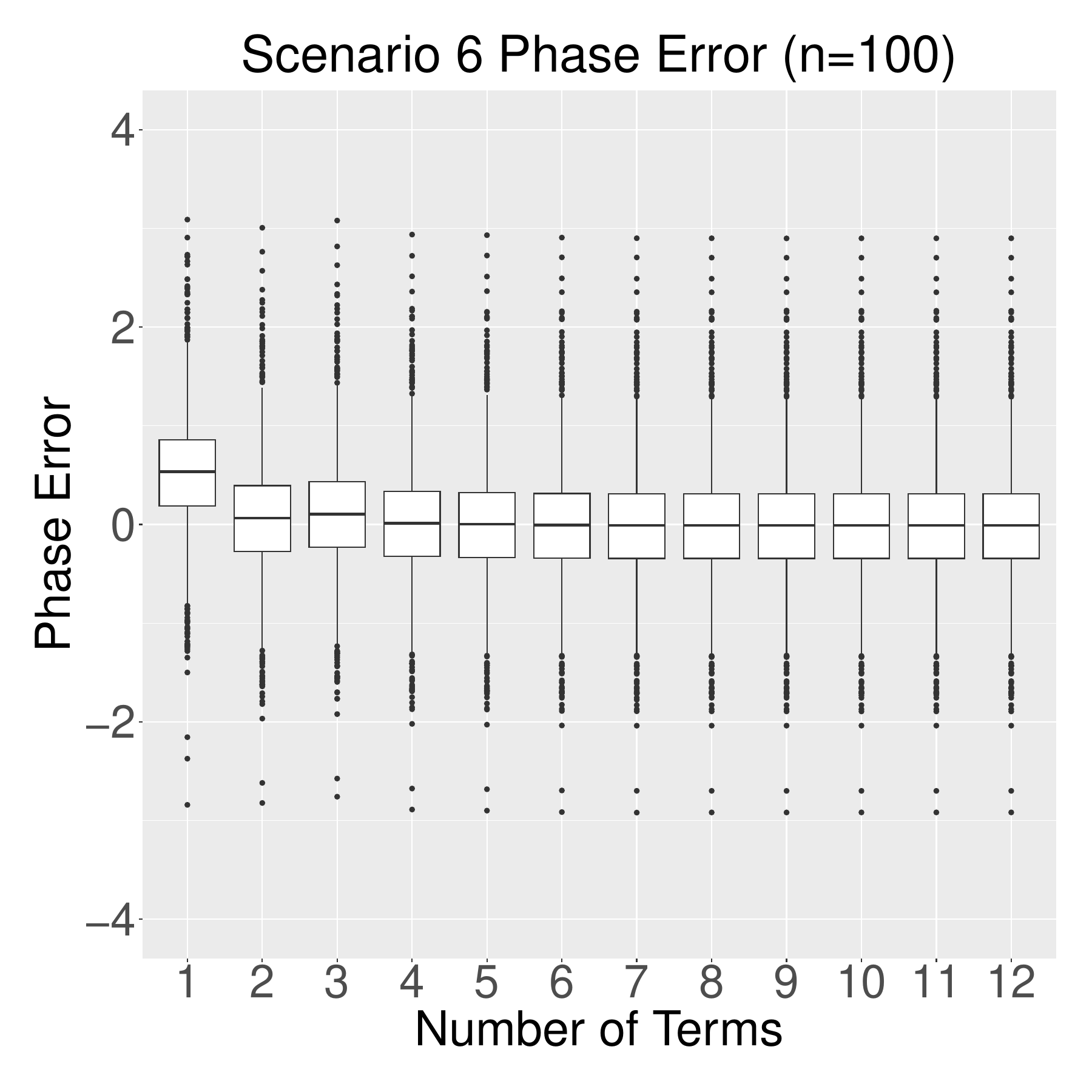}
\includegraphics[scale=0.14]{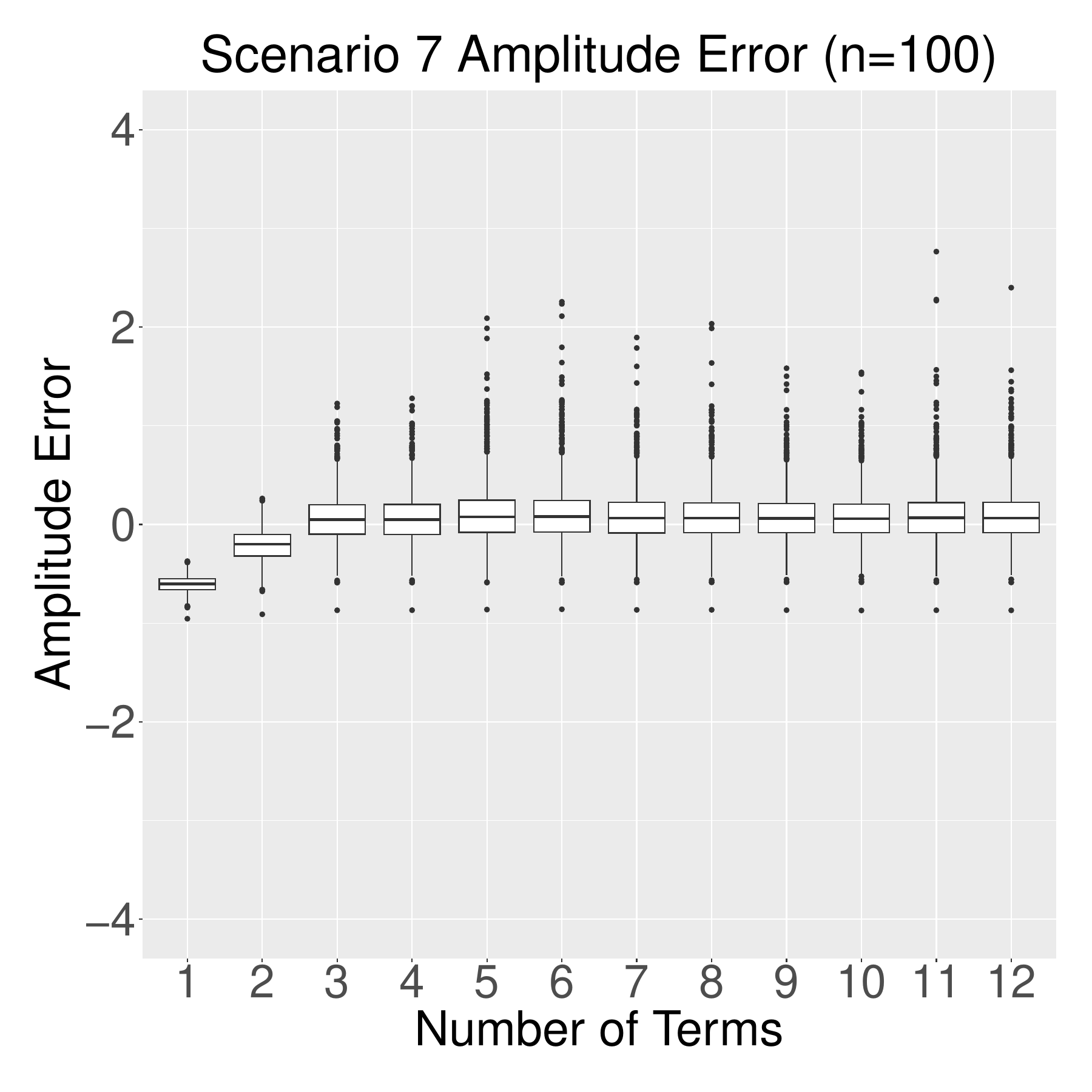}
\includegraphics[scale=0.14]{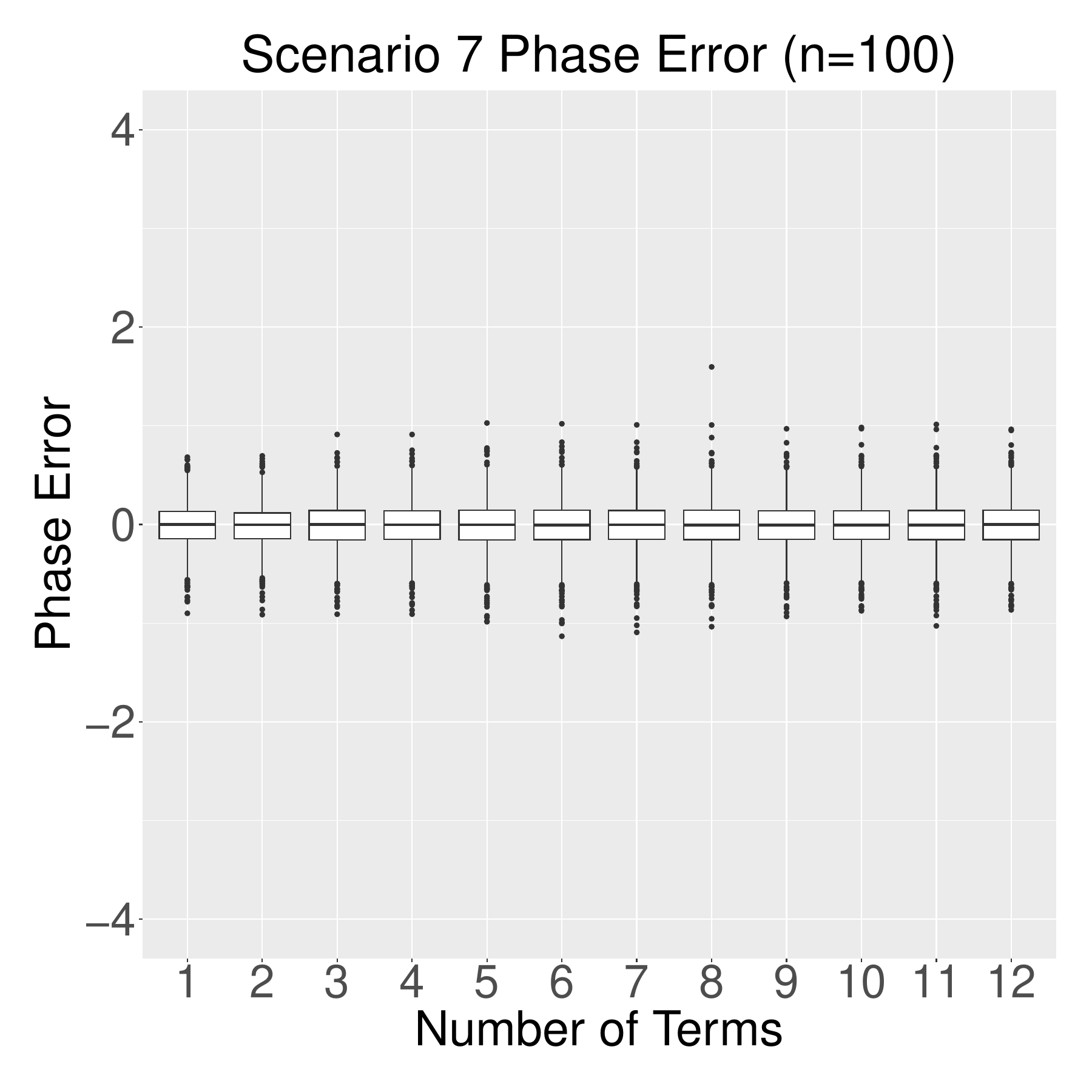}
\includegraphics[scale=0.14]{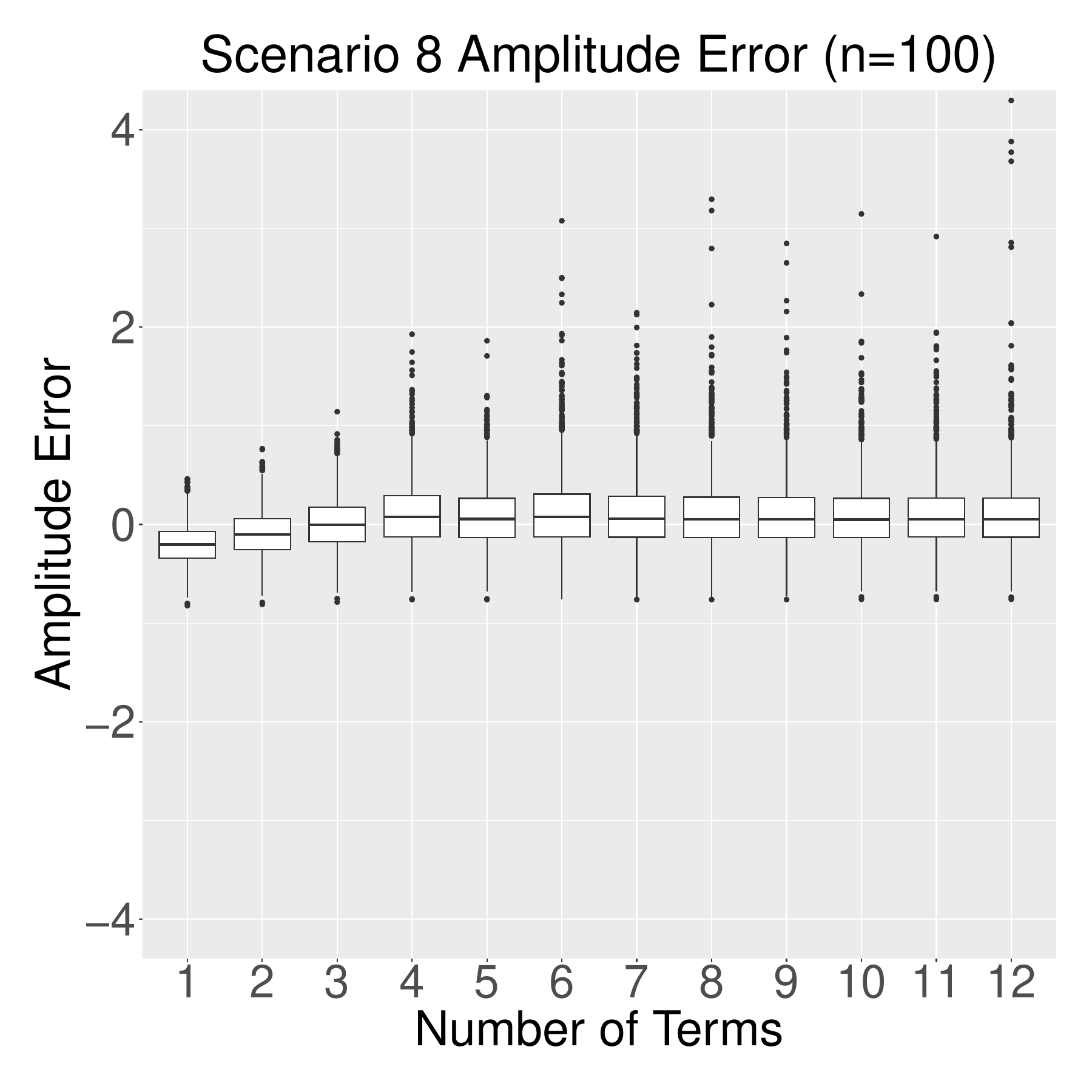}
\includegraphics[scale=0.14]{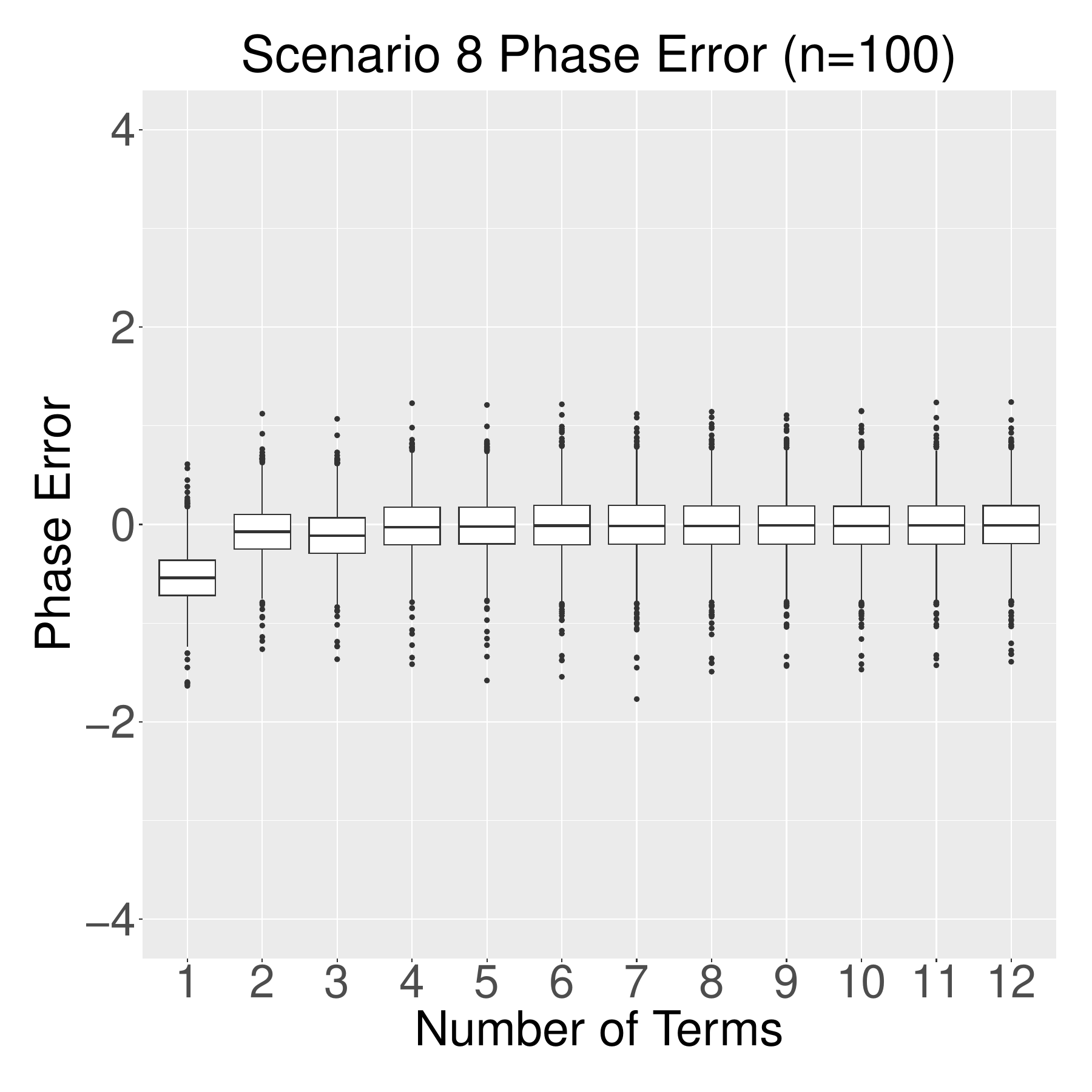}
\caption{Box plots of the error in amplitude estimation, or $(\hat{A}-A^*)/A^*$, and phase-shift estimation, or $\hat{\eta}-\eta^*$, for the cosinor regression model from (\ref{eq:3}) as the number of terms from the corrected score function in Theorem \ref{thm1} varies from 1 to 12 ($m=0,\ldots,11$) and the sample size is $n=100$. When the number of terms equals one ($m=0$), the resulting estimates are the same as those obtained from a score function that does not account for measurement error.}
    \label{fig:terms}
\end{figure*}

Given that Framework 1 uses the first three terms of the corrected score function from Theorem \ref{thm1}, Figure \ref{fig:terms} provides an assessment of parameter estimation with (\ref{eq:cor_cond}) when a different number of terms are used and the sample size $n=100$. Specifically, Figure \ref{fig:terms} indicates that as the number of terms from the infinite summation increases, estimation with the corrected score function can produce inflated amplitude estimates. Corresponding assessments when the sample size is $n=400$ and $n=1,600$ are provided in Section 1 of the supplementary material, which demonstrate that this inflation phenomenon disappears as sample size increases.

\section{Illustrations with Circadian Transcriptomic Data} \label{sec:4}

\subsection{Overview of data sets} \label{sec:4.1}

Three circadian transcriptomic data sets are employed for illustration. The ``Archer data set'' is microarray data from a study that investigates how gene expression changes before and after a four day (96 hour) sleep desynchrony protocol in 22 people \citep{Archer2014}. The sleep desynchrony protocol involved each individual adhering to a 28 hour wake-sleep cycle, or they were required to stay awake for a 20 hour period followed by an eight hour sleep period. For each individual, two sets of blood samples were collected to generate transcriptomic data. The first set was collected once every four hours from each individual over a 24 hour period before the sleep desynchrony protocol (control group samples). The second set was collected once every four hours over a 24 hour period after the protocol (experimental group samples). Two additional sets of blood samples were drawn from each individual to determine their DLMO times before and after the experiment. Each set of blood samples for determining DLMO times was acquired during the same 24 hour periods as the blood samples for deriving transcriptomic data. However, the blood samples for determining DLMO times were collected at one to two hour intervals \citep{hasan2012}. 

The ``Braun data set'' is next-generation sequencing data from a single control cohort \citep{Braun2018}. Blood samples were collected from 11 people who had similar sleep schedules, were medically stable with no mental health conditions, and were between 18 and 60 years of age. Sample collection occurred once every two hours over the course of a 29 hour period. At each time point of sample collection, two samples were obtained: one to derive transcriptomic data, and another to determine each individual's DLMO time. 

The ``M\"{o}ller-Levet data set'' is microarray data from a study investigating how insufficient sleep affects gene expression in 26 people \citep{MllerLevet2013}. This study obtained two sets of blood samples to generate transcriptomic data. The first set was procured following a ten day period during which each individual was allowed up to ten hours of sleep per night (control group samples). The second set was obtained after a ten day period during which each individual was restricted to six hours of sleep per night (experimental group samples). Both sets consist of samples that were collected from each individual once every three hours over a 30 hour period. Additionally, saliva samples were collected hourly during these 30 hour periods to determine the DLMO time for each individual after each ten day period.

For illustration, seven different sample populations are obtained from these three data sets. With the Archer data set, three sample populations were formed: one from the control group samples, another from the experimental group samples, and a third from both groups combined. With the Braun data set, a single sample population was formed. Finally, with the M\"{o}ller-Levet data set, three sample populations were formed: one from control group samples, another from experimental group samples, and a third from both groups combined. 

Each data set was processed and made publicly available \citep{Braun2018, Braun2019}. Every processed data set consists of 7,615 genes, 50 of which are gold standard circadian genes that have been recognized as displaying oscillatory behavior across multiple populations \citep{Mei2020}. For this study, two additional processing steps were performed with these sample populations. The first processing step removed samples where internal circadian time (ICT, or the true covariate $X^*$) could not be determined. After this processing step, the M\"{o}ller-Levet data set consisted of $n=181$ samples from the control group, and $n=174$ samples from the experimental group ($n=355$ for both groups combined); the Archer data set consisted of $n=127$ samples from the control group, and $n=131$ samples from the experimental group ($n=258$ for both groups combined); and the Braun data set consisted of $n=153$ samples. The second processing step eliminated genes with missing expression data. This processing step affected the Archer data set, reducing it to 4,475 genes for the control group (28 gold standard circadian genes), 4,599 genes for the experimental group (31 gold standard circadian genes), and 3,689 genes for both groups combined (23 gold standard circadian genes). A summary of each sample population before and after processing is provided in Table \ref{tab:data_sum}.

\subsection{Overview of DLMO time data sets} \label{sec:4.2}

\begin{figure*}[!h]
\centering
\includegraphics[scale=0.14]{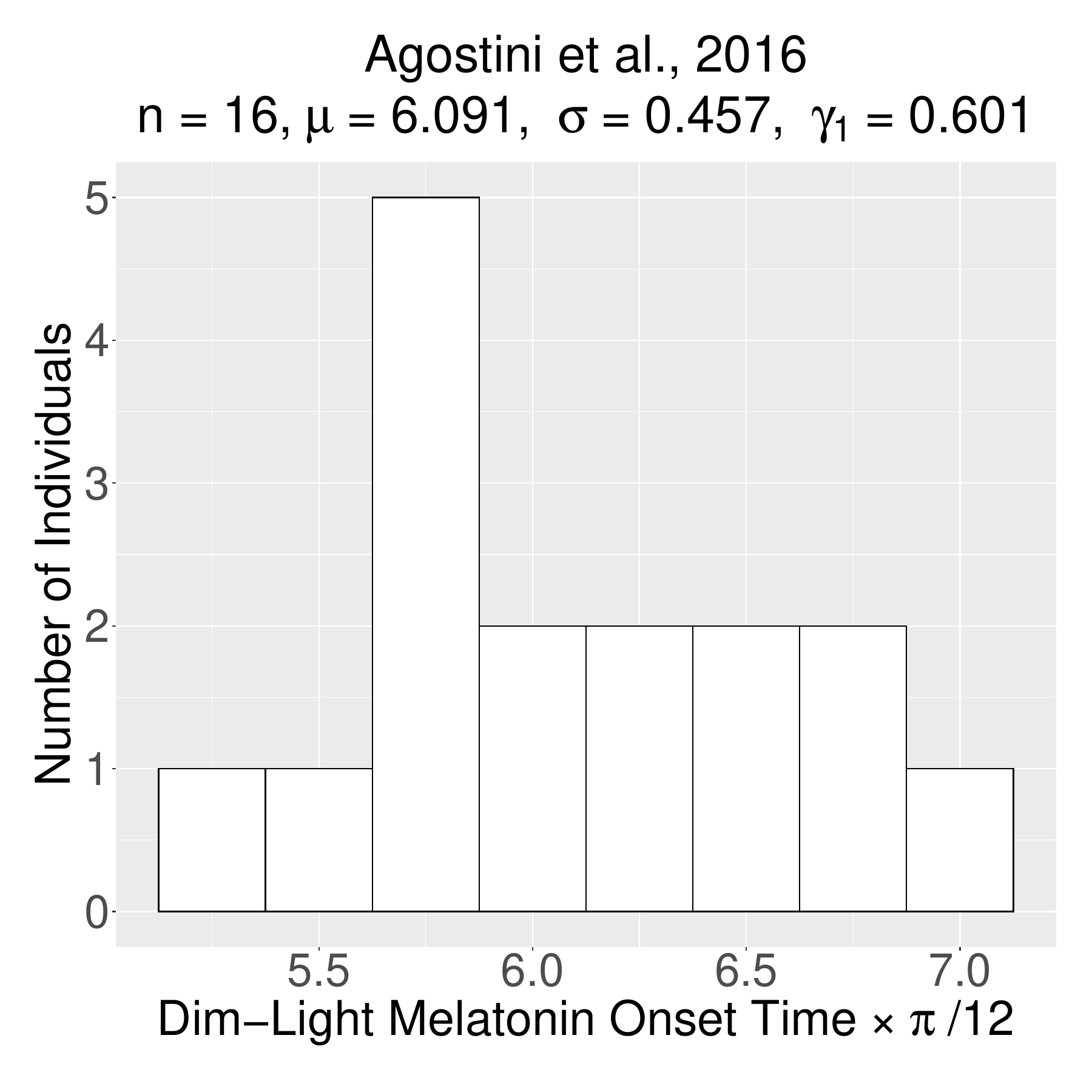}
\includegraphics[scale=0.14]{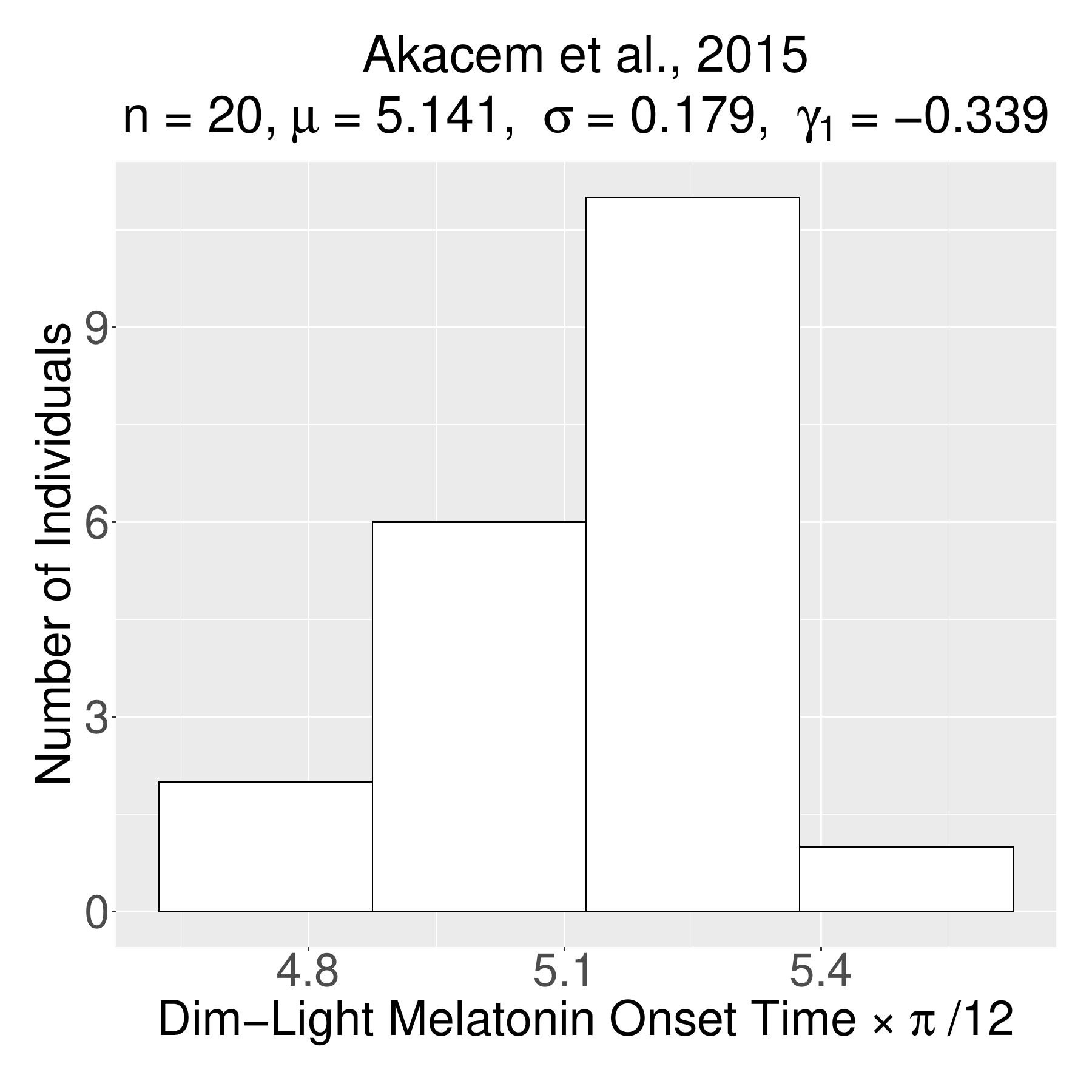}
\includegraphics[scale=0.14]{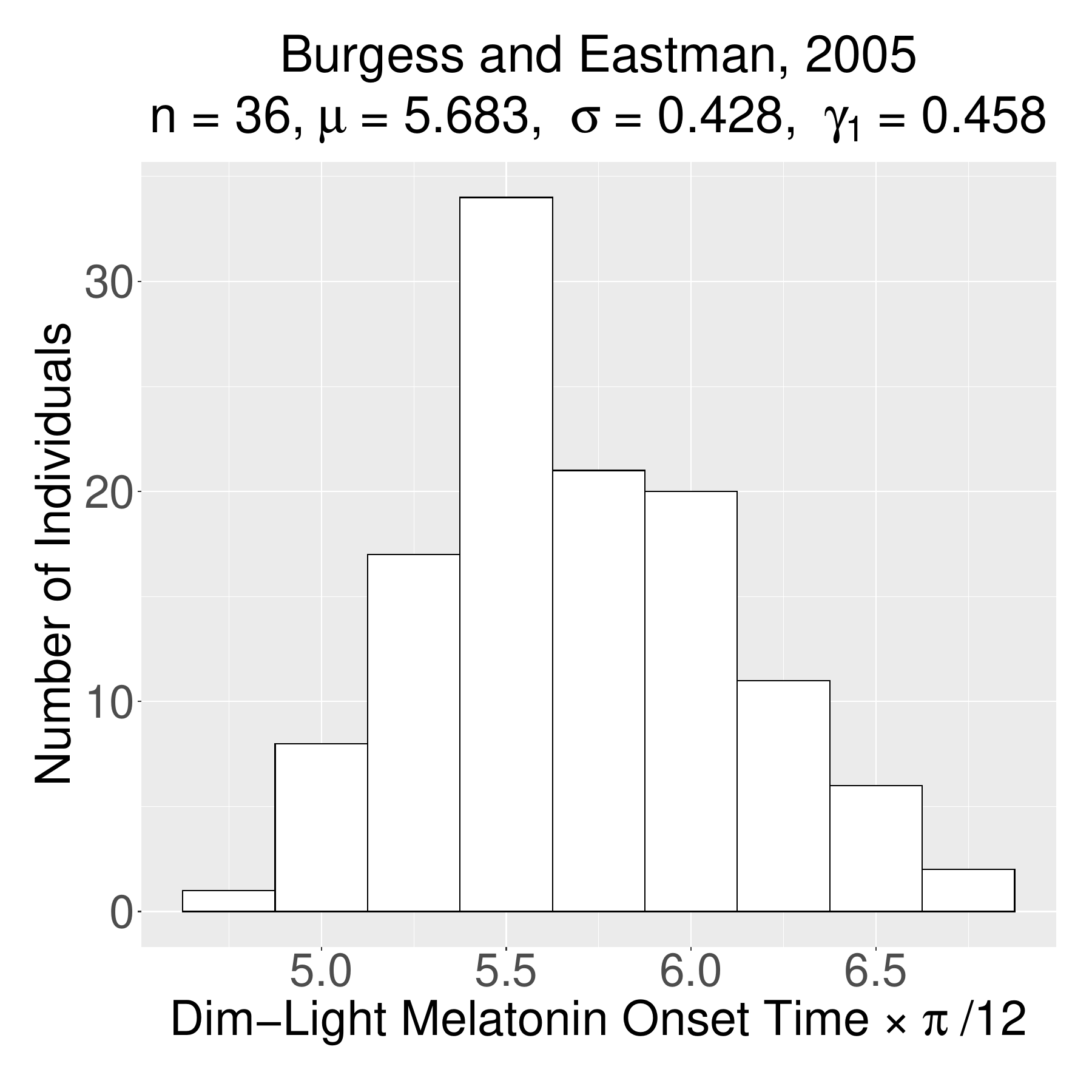}
\includegraphics[scale=0.14]{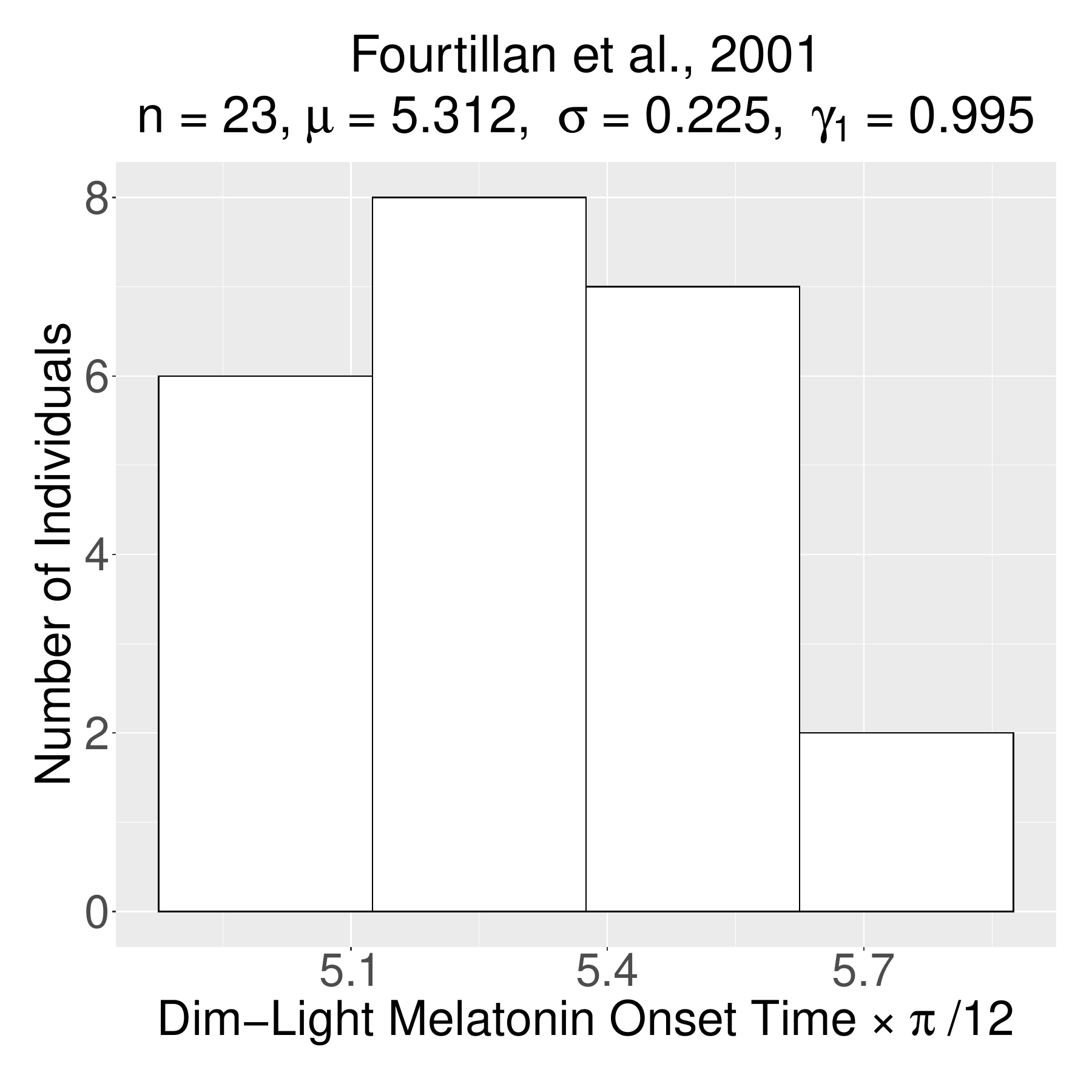}
\includegraphics[scale=0.14]{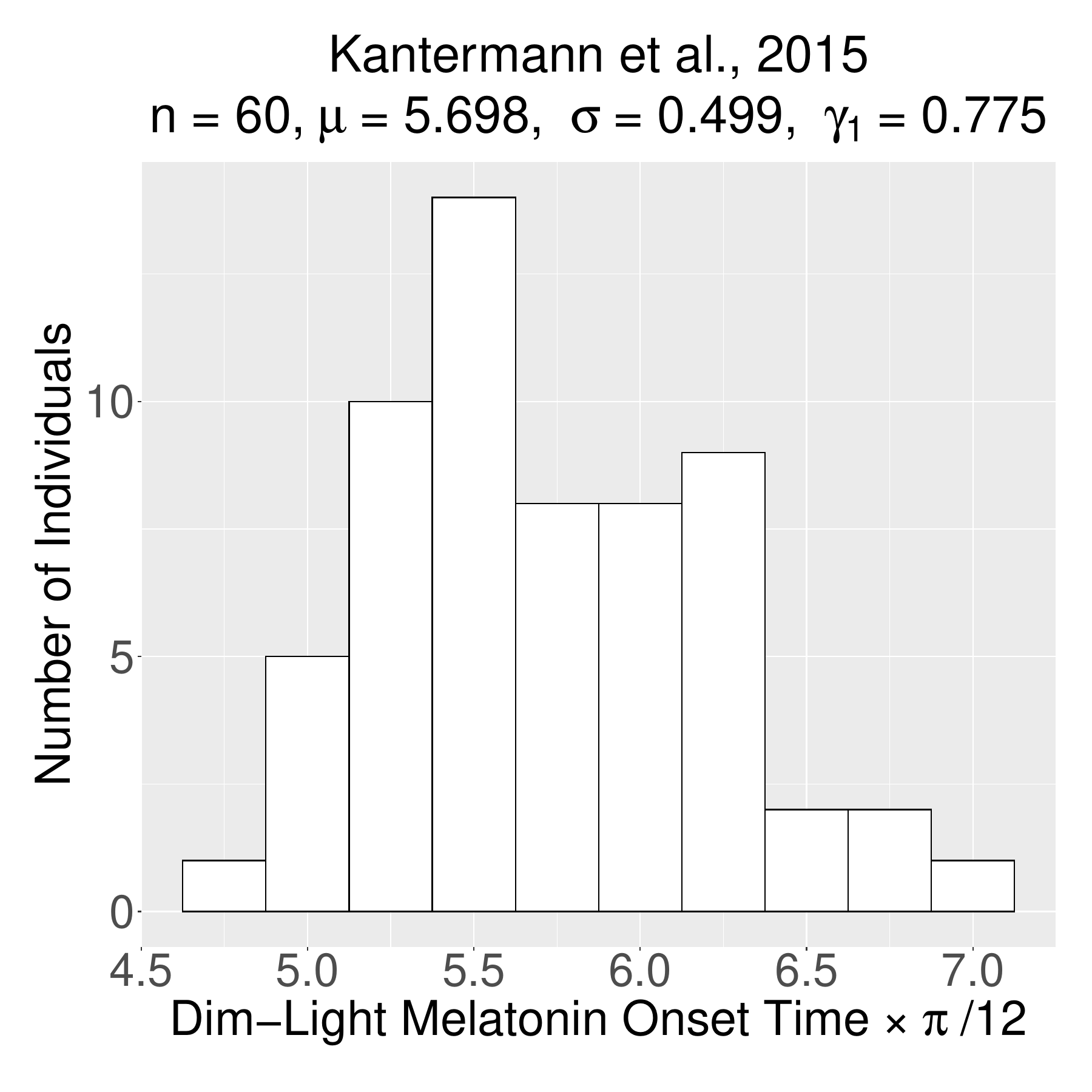}
\includegraphics[scale=0.14]{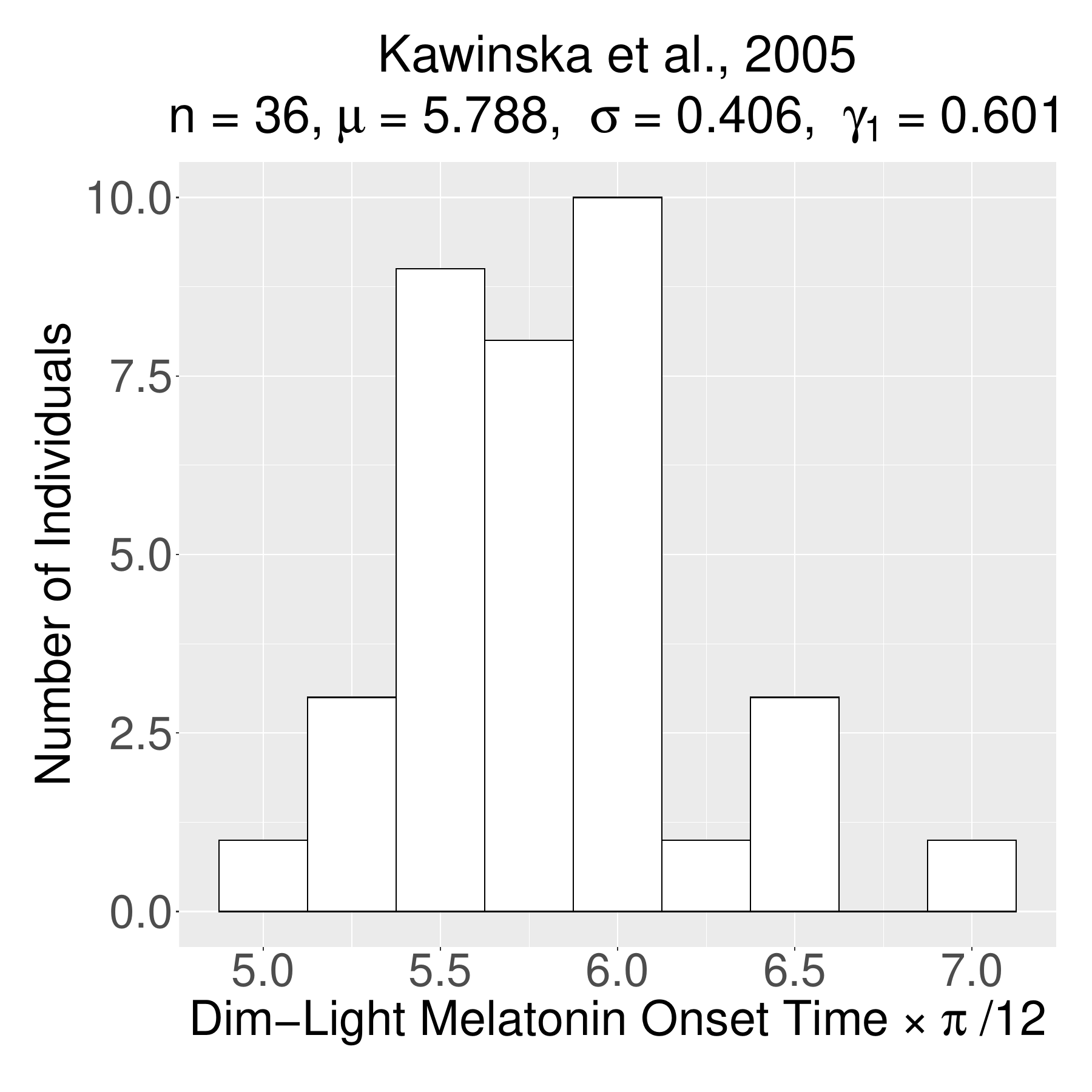}
\includegraphics[scale=0.14]{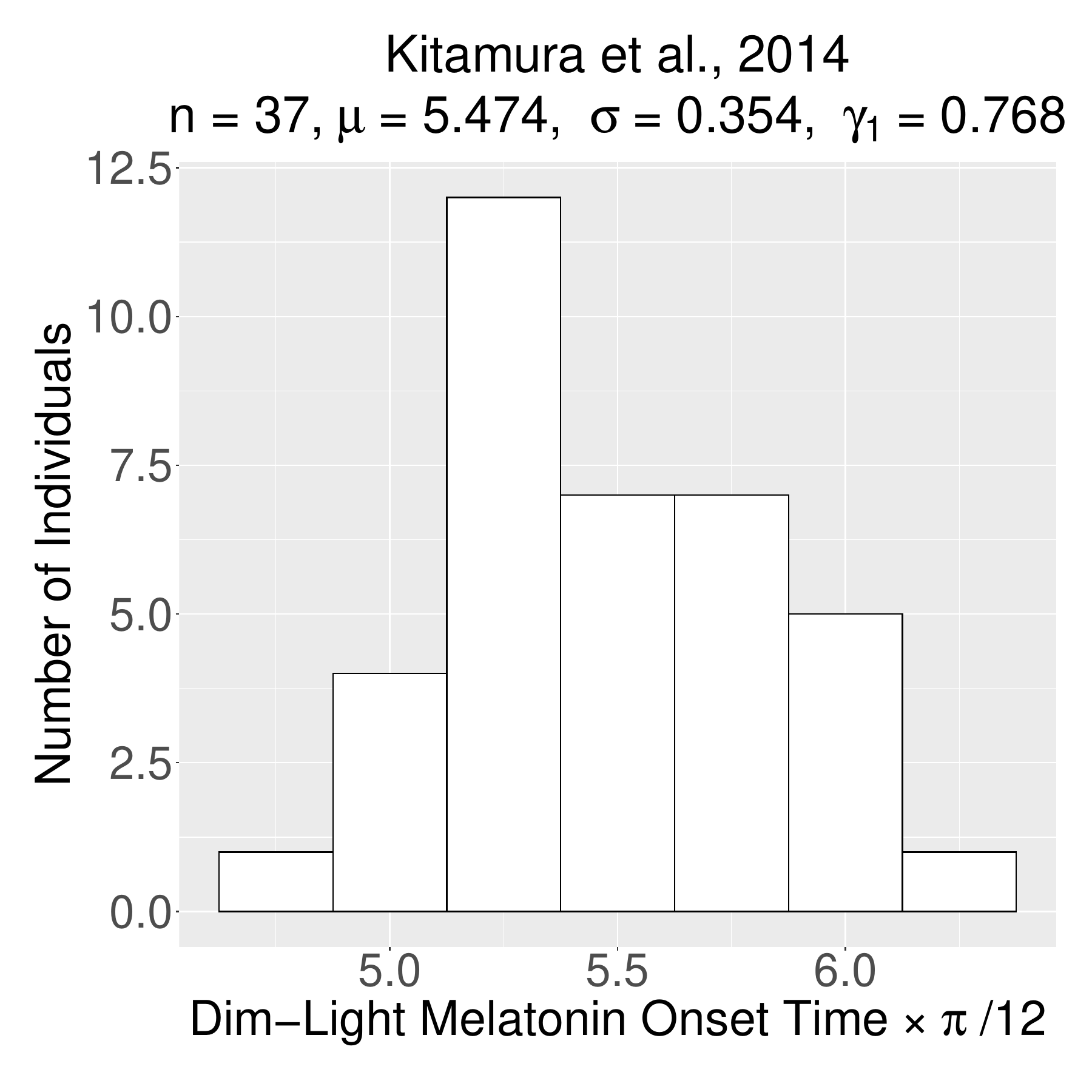}
\includegraphics[scale=0.14]{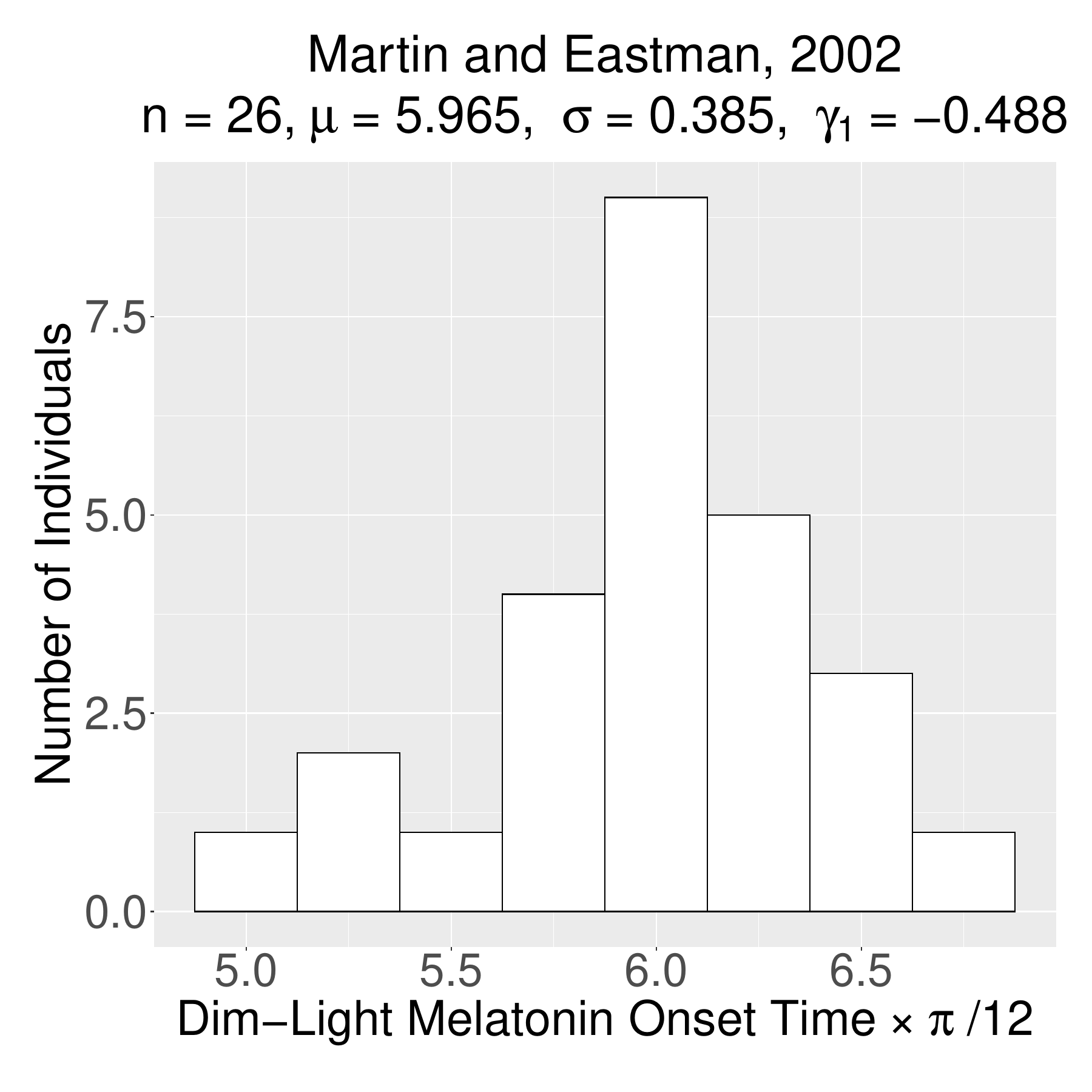}
\includegraphics[scale=0.14]{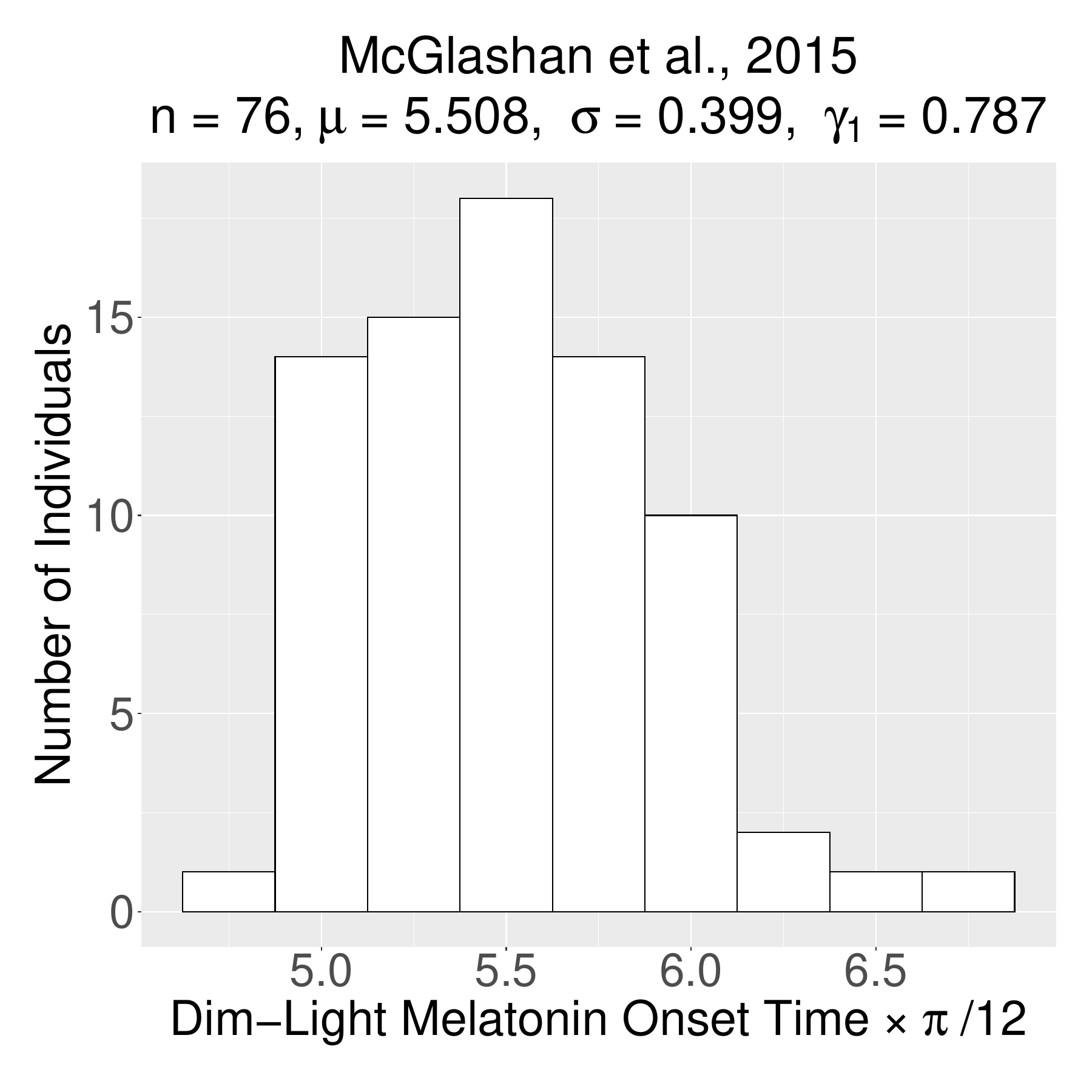}
\includegraphics[scale=0.14]{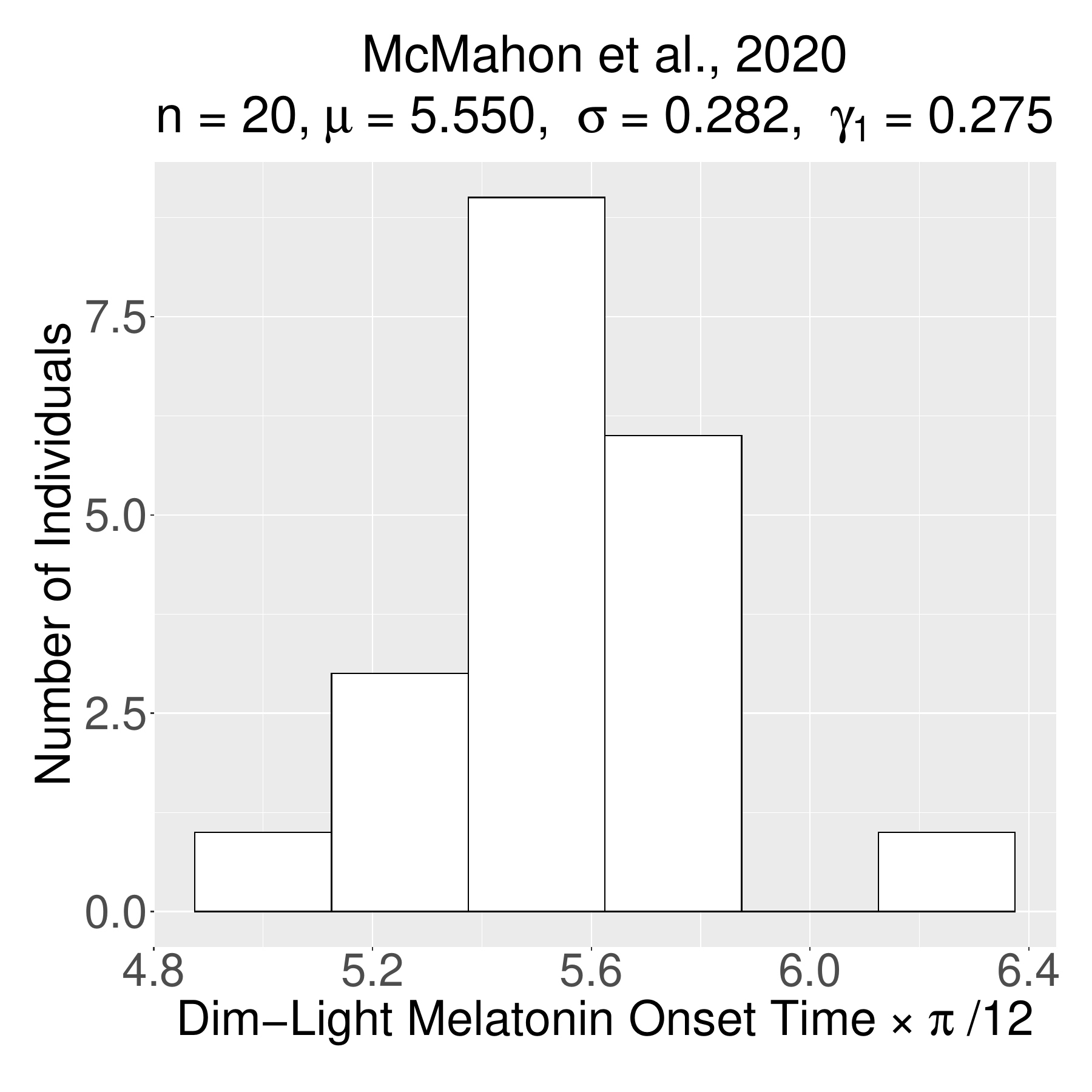}
\includegraphics[scale=0.14]{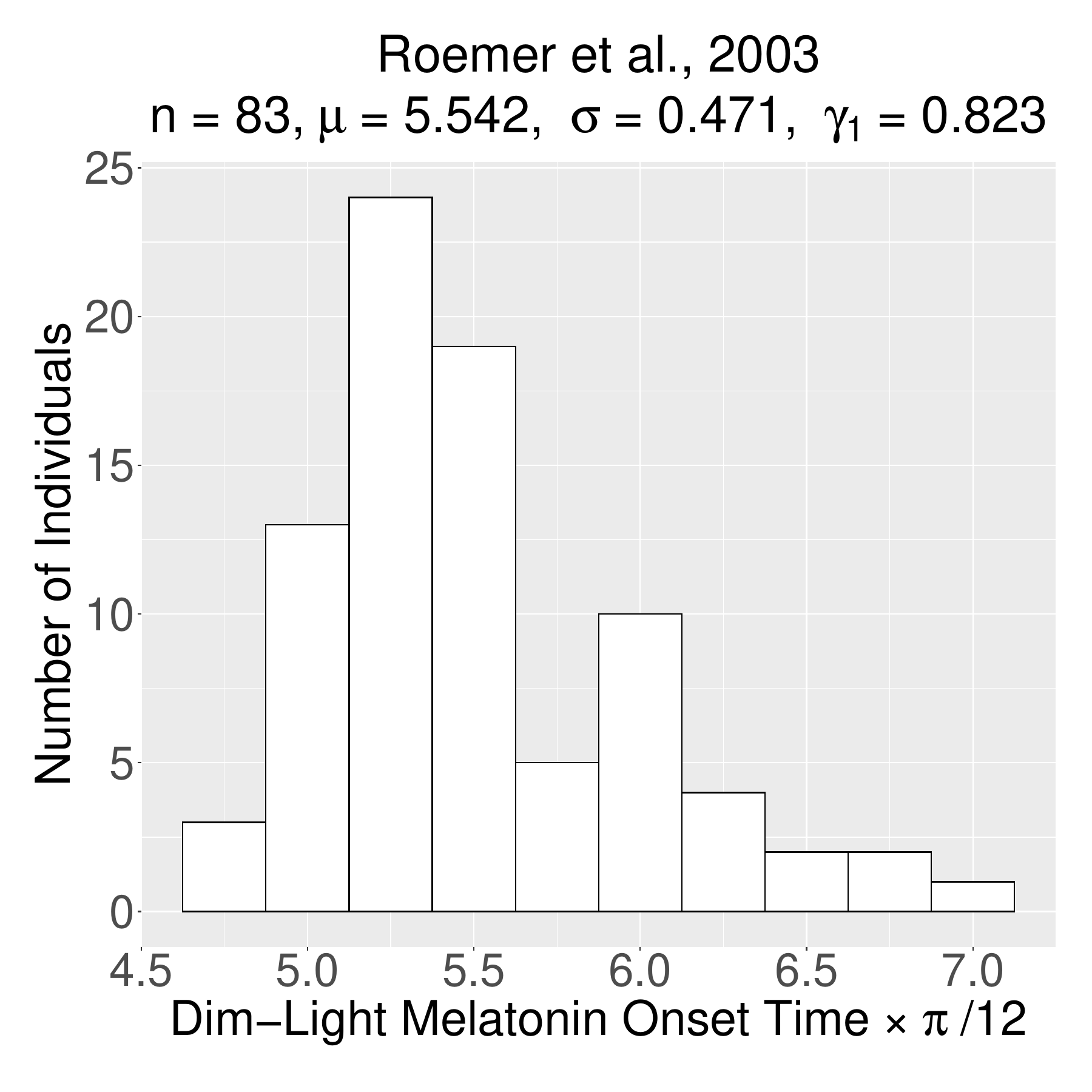}
\includegraphics[scale=0.14]{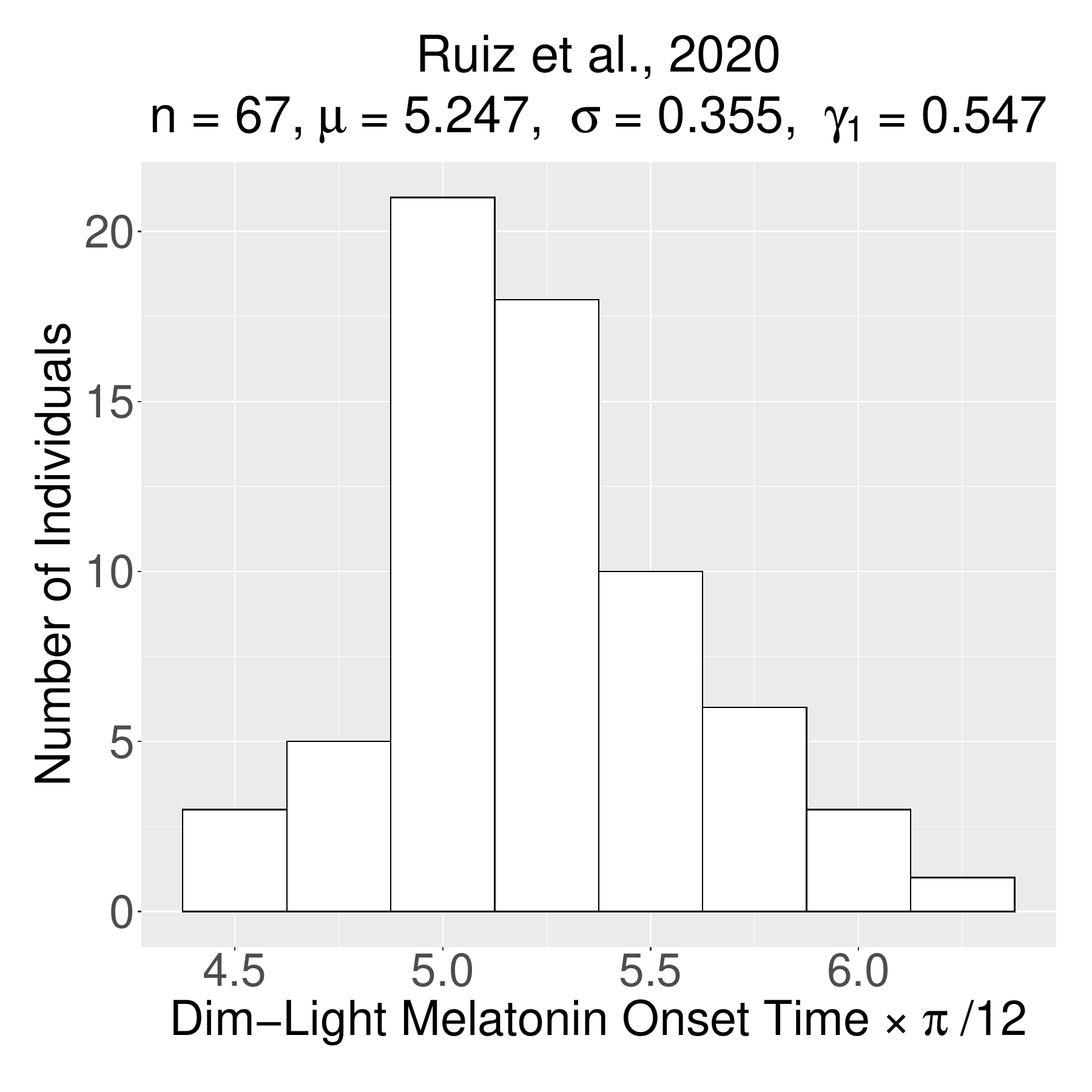}
\includegraphics[scale=0.14]{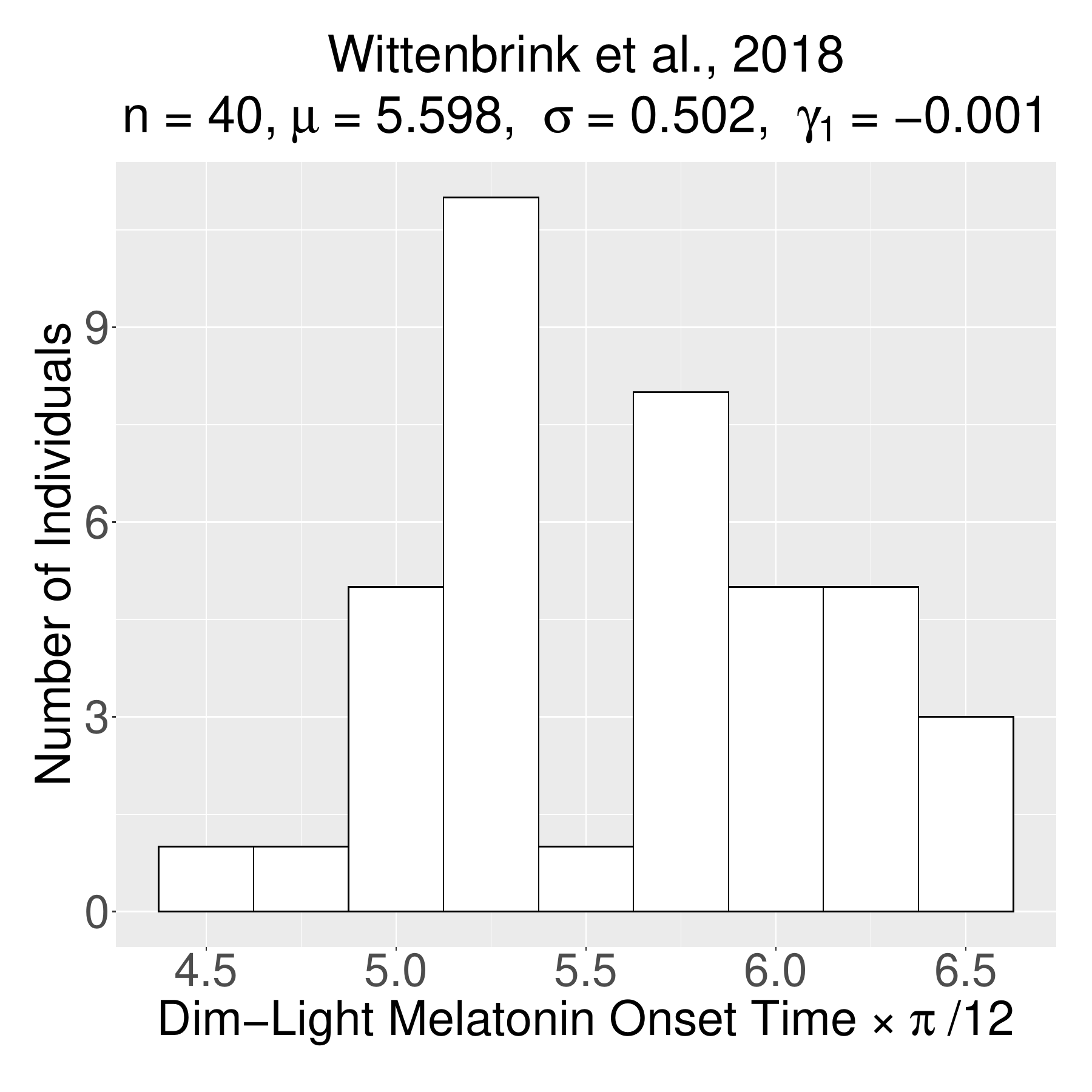}
\includegraphics[scale=0.14]{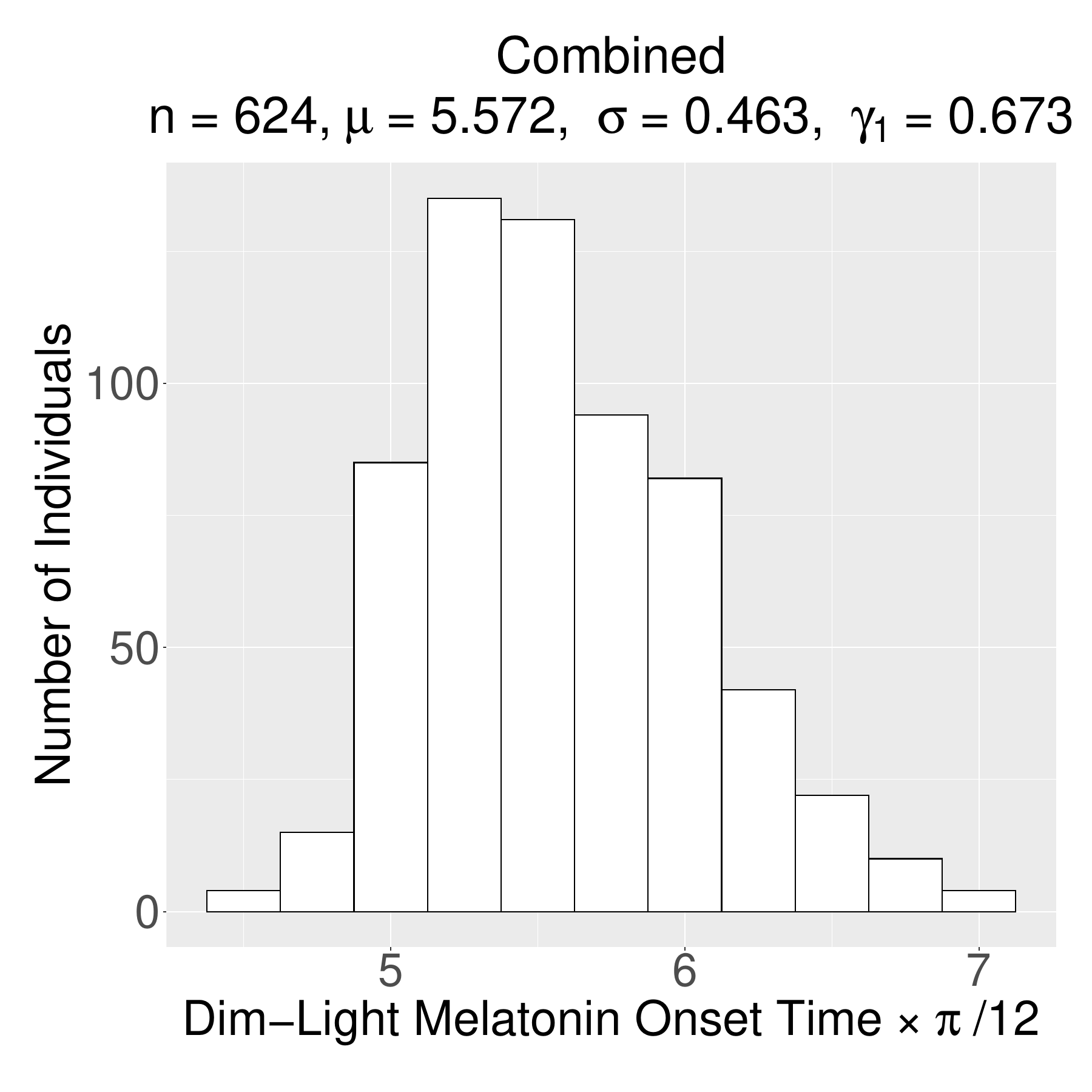}
\caption{Histograms of Zeitgeber times for melatonin onset under dim-light conditions (DLMO times) recorded in 13 different studies. The title of each plot reports the study, sample size ($n$), mean DLMO time ($\mu$), standard deviation of DLMO times ($\sigma$), and skewness of DLMO times ($\gamma_1$) when modeled with a skew normal distribution.}
    \label{fig:13_data}
\end{figure*}

The corrected score function from Theorem \ref{thm1} requires an investigator to specify a numeric value for each $\kappa^{(m)}(0)$, which relates to the moments of the probability distribution generating DLMO times. In this paper, 13 additional DLMO time data sets are obtained from published studies \citep{Kennaway2023}. A summary of the empirical DLMO time distributions for each additional data set is provided in Figure \ref{fig:13_data}. This summary indicates that most DLMO time distributions are asymmetric with a non-zero mean. Further, none of these distributions have a standard deviation that exceeds 2 hours. Table \ref{tab:kappas} provides estimates of each $\kappa^{(m)}(0)$, $m=1,\ldots,8$ obtained from these 13 additional data sets, with the smallest value for each $\kappa^{(m)}(0)$ denoted in bold. Each $\kappa^{(m)}(0)$ is computed from the product of DLMO time and $\pi/12$, subtracted by $2\pi$.

\subsection{Illustration setup} \label{sec:4.3}

The illustrations follow a similar setup to that of the simulation study in Section \ref{sec:3}. First, two estimators are compared on circadian transcriptomic data: one that satisfies (\ref{eq:cor_cond}) given ZT data and expression data from a single gene (Framework 1 from Section \ref{sec:3}), and another that satisfies (\ref{eq:emp_grad}) given ZT data and expression data from a single gene (Framework 2 from Section \ref{sec:3}). For Framework 1, each value of $\kappa^{(m)}(0)$ for the corrected score function in Theorem \ref{thm1} is now defined as the quantity that is smallest in magnitude across all 13 data sets in Table \ref{tab:kappas}. The selection of these values reflects a conservative assumption that the differences between $X^{\dagger}$ and $X^*$ are modest, as setting each $\kappa^{(m)}(0) = 0$ for all $m>0$ is equivalent to identifying the parameter vector that satisfies (\ref{eq:emp_grad}). The selection of these numeric values for each $\kappa^{(m)}(0)$ also enables application of the corrected score function on each of the seven sample populations without making further assumptions. It is noted that estimation and inference with Framework 1 will again use the first three terms ($m=0,1,2$) of the corrected score function.

The assessment of these two frameworks also mirrors the assessment conducted in the simulation study from Section \ref{sec:3}, which considered accuracy in parameter estimation for the model from (\ref{eq:3}) and score test statistics scaled by sample size from (\ref{eq:st_t}). However, there are two notable differences in this assessment. First, the true values of these parameters and score test statistics are unknown. To address this, the true quantities are replaced with empirical quantities computed from ICT data. Second, each data set comprises expression levels from multiple genes, and each gene displays different oscillatory behavior over time. In other words, for each recorded $X_j^{\dagger}$, there is a corresponding $Y_j$ recorded for each gene, and the parameter vector estimate obtained for one gene can differ from the parameter vector estimates obtained for every other gene. To address the availability of multiple genes, a linear model-based assessment is proposed. In this assessment, a linear model is specified where a quantity (estimate of a parameter or score test statistic) obtained from ICT and gene expression (ICT-based data) is specified as the response variable, and a corresponding quantity obtained from ZT and gene expression (ZT-based data) with a framework is specified as the covariate. An intercept term is not specified for these linear models, which results in a single regression parameter estimate $\beta$. This assessment employs $\beta$ to evaluate the relationship between quantities obtained from ICT-based data and quantities obtained from ZT-based data with a framework. Specifically, a value of $\beta=1$ indicates a linear relationship between the quantities obtained from ICT-based data and a framework using ZT-based data. A value of $\beta > 1$ indicates that quantities obtained from a framework using ZT-based data are attenuated relative to using ICT-based data, and a value of $\beta < 1$ that quantities obtained from a framework using ZT-based data are inflated relative to using ICT-based data. To complement assessment of each $\beta$ obtained, the coefficient of determination ($R^2$) is also computed, where higher $R^2$ values signify greater precision in the linear model fit. It is noted that a stronger performing framework would obtain $\beta$ closer to one and a large $R^2$ value.

To provide a concrete example for this assessment setup, the control group sample population from the Archer data set has 4,475 genes available, which would result in 4,475 cosinor regression models estimated under a framework. When evaluating the amplitude estimates obtained from ZT-based data with a framework ($\hat{A}$) relative to those obtained from ICT-based data ($A^*$), a linear model is specified as $A^{*}=\beta\hat{A}$, and this linear model is estimated by minimizing squared loss. Both the parameter estimate $\beta$ and the coefficient of determination $R^2$ computed from this linear model are subsequently reported. It is emphasized that the phase-shift $\eta^*$ is a circular quantity restricted to the interval $[-\pi, \pi)$. To account for this, when the absolute difference $|\hat{\eta}-\eta^*| > \pi$, $\hat{\eta}$ is mapped to $\hat{\eta}+2\pi$ prior to regression when $\eta^* > 0$, and $\hat{\eta}$ is mapped to $\hat{\eta}-2\pi$ prior to regression when $\eta^* < 0$.

\subsection{Results} \label{sec:4.4}

Table \ref{tab:app1} presents the results from linear model-based assessments on each of the seven sample populations. These results demonstrate that the linear model parameter estimate $\beta$ is consistently closer to one for Framework 1, which indicates that it consistently mitigates bias in parameter estimation and inference. Further, Framework 2 consistently obtains a value of $\beta > 1$ for amplitude estimates and score test statistics, which indicates that there is an attenuation bias in amplitude parameter estimation and a loss of power in hypothesis testing. For amplitude estimation, Framework 1 outperforms Framework 2 on five of the seven sample populations. For phase-shift estimation, Framework 1 outperforms Framework 2 on six of the seven sample populations (both frameworks obtain the same performance on the remaining sample population). Finally, for score test statistic calculation, Framework 1 outperforms Framework 2 on five of the seven sample populations. Framework 1 also consistently achieves $R^2$ values exceeding 0.9 in each assessment. It is emphasized that both frameworks obtain values of $\beta=1$ and $R^2=1$ for estimation accuracy of the intercept $\theta^*_0$, which has been omitted from Table \ref{tab:app1}.

Table \ref{tab:app2} displays the results for each of the seven sample populations when linear models are estimated with quantities obtained from the gold standard circadian genes available. Framework 1 consistently outperforms Framework 2 on this subset of genes, and Framework 2 again consistently displays attenuation bias in amplitude estimation as well as a loss of statistical power in hypothesis testing. When using $\beta$ to assess accuracy in amplitude and phase-shift parameter estimation (\ref{eq:3}), Framework 1 is more accurate than Framework 2 in amplitude estimation for five out of seven sample populations and in phase-shift estimation for four out of seven sample populations (both frameworks obtain the same performance on one of the three remaining sample population). Framework 1 also outperforms Framework 2 in the calculation of score test statistics for six out of seven sample populations. Both frameworks again obtain values of $\beta=1$ and $R^2=1$ for estimation accuracy of the intercept $\theta^*_0$, which has been omitted from Table \ref{tab:app2}. Scatter plots that display the fit of each linear model relative to the quantities obtained from a framework and a corresponding quantity obtained from ICT-based data are provided in Section 2 of the supplementary material.

\section{Discussion} \label{sec:5}

In this paper, we propose a corrected score function framework for regression modelling of circadian transcriptomic study data. The development of this framework is motivated by each individual in a study having an internal circadian time (ICT) that could be offset relative to the 24 hour day-night cycle time (Zeitgeber time or ZT) \citep{Lewy1999, Ruiz2020, Wittenbrink2018, Wright2013}. Many corrected score function frameworks make the assumption that the probability distribution generating offsets is symmetric with a mean of zero \citep{Cai2021, Hong2003, Nakamura1990, Nakamura1992, Stefanski1987}. The proposed framework removes this assumption to accommodate asymmetric probability distributions with a non-zero mean, which was validated with both simulated and real data. It is emphasized that application of this corrected score function framework requires an investigator to specify a numeric value for each $\kappa^{(m)}(0)$ Theorem \ref{thm1}. We acquire these values from 13 studies that provide data on the variation of ZT at which melatonin onset occurs under dim-light conditions (DLMO time) for different cohorts \citep{Kennaway2023}. DLMO time serves as a gold standard marker for this offset \citep{Lewy1999, Ruiz2020, Wittenbrink2018}.

It is crucial to address the limitations of this study. One limitation is that the accuracy of the corrected score function in Theorem \ref{thm1} depends on precise estimates of each $m$-th order derivative for $\kappa(t)$ evaluated at zero, or $\kappa^{(m)}(0)$. Given that estimates and inferences were sometimes more accurate with a score function that did not account for measurement error, we recommend that the specified numeric values for each $\kappa^{(m)}(0)$ are no larger than the smallest values identified in Table \ref{tab:kappas} when there is not a strong understanding of how circadian transcriptomic data are mismeasured. Specification of small values would ensure that the corrected score function framework mitigates bias in estimation and inference relative to using a score function that does not account for measurement error and satisfies the absolute convergence criteria in Theorem \ref{thm1}. A second limitation is that the corrected score function in Theorem \ref{thm1} is derived under the assumption of classical measurement error. This assumption could be invalid. However, the simulation study results in Section \ref{sec:3} indicates that corrected score function can empirically accommodate Berkson measurement error, which is another source of measurement error in data commonly assumed \citep{Berkson1950}.

The work presented in this study opens up avenues for future research. First, one could consider applying recent non-parametric methods for mismeasured data \citep{Delaigle2016, DiMarzio2021, Dimarzio2023, Nghiem2018, Nghiem2020} to circadian transcriptomic data. Second, one could consider using additional data obtained from a cohort, such as surveys, to produce each $\kappa^{(m)}(0)$ \citep{Kantermann2015, kitamura, Martin, Ruiz2020}.

\section*{Acknowledgements}
The authors thank Professor David Kennaway at the University of Adelaide for identifying the 13 DLMO time studies in Section \ref{sec:4.2}. This work was performed under the auspices of the U.S. Department of Energy by Lawrence Livermore National Laboratory under Contract DE-AC52-07NA27344.

\section*{Conflict of interest}
The authors declare no potential conflict of interests.

\section*{DATA AVAILABILITY STATEMENT}
The authors have made each of the 13 DLMO time data sets from Section \ref{sec:4.2} as well as code scripts to reproduce Tables \ref{tab:app1} and \ref{tab:app2} available at \textcolor{blue}{\href{https://bitbucket.org/michaelgorczyca/eiv_dlmo/}{https://bitbucket.org/michaelgorczyca/eiv\_dlmo/}}.

\section*{Supporting information}
The supplementary material includes simulation results for larger sample sizes from Section \ref{sec:3}, as well as with scatter plots of linear model fits for the data illustrations from Section \ref{sec:4.4}. 

\bibliographystyle{apalike}
\bibliography{bibliography}

\clearpage

\newpage

\begin{table*}[!b]
	\caption{Simulation study results for both frameworks. Framework 1 uses the corrected score function from Theorem \ref{thm1} for obtaining each quantity. Framework 2 uses a score function that does not account for measurement error. ``Bias'' denotes the difference between the estimated and true quantity, and ``Abs. Err.'' denotes the corresponding absolute error. The mean difference and mean absolute difference are reported, with their corresponding standard deviations in parentheses. Mean difference and mean absolute difference values closer to zero denoted in bold.} \label{tab:sim1}
 \centering
		\resizebox{1.0\textwidth}{!}{
  \begin{tabular}{|c|c|c|c|c|c|c|c|c|c|}
			\hline
   \multirow{3}{*}{Scenario} & \multirow{3}{*}{Framework} & \multicolumn{8}{c|}{ $n=100$ Samples} \\
   \cline{3-10} 
			 &  & \multicolumn{2}{c|}{$\hat{\theta}_0-\theta^*_0$} & \multicolumn{2}{c|}{$(\hat{A}-A^*)/A^*$} & \multicolumn{2}{c|}{$\hat{\eta}-\eta^*$} & \multicolumn{2}{c|}{$(\tau_{\text{Score}}-\tau^*_{\text{Score}})/n$} \\
   \cline{3-10}
   & & Bias & Abs. Err.& Bias & Abs. Err.& Bias & Abs. Err.& Bias & Abs. Err.  \\
   \hline
        \multirow{2}{*}{1}    & 1 & \textbf{-0.002 (0.081)} & 0.052 (0.050) & \textbf{0.019 (0.152)} & \textbf{0.092 (0.102)} & \textbf{0.000 (0.144)} & 0.095 (0.090) & \textbf{0.003 (0.106)} & \textbf{0.070 (0.066)} \\
                              & 2 & \textbf{-0.002 (0.062)} & \textbf{0.042 (0.038)} & -0.221 (0.086) & 0.221 (0.085) & 0.002 (0.122) & \textbf{0.085 (0.072)} & -0.199 (0.065) & 0.199 (0.065) \\
         \hline 
         \multirow{2}{*}{2}   & 1 & -0.002 (0.135) & 0.091 (0.082) & \textbf{0.083 (0.452)} & 0.302 (0.289) & \textbf{-0.136 (0.601)} & \textbf{0.372 (0.401)} & \textbf{0.018 (0.129)} & 0.072 (0.096) \\
                              & 2 & \textbf{-0.001 (0.133)} & \textbf{0.089 (0.080)} & -0.131 (0.351) & \textbf{0.257 (0.213)} & -0.551 (0.604) & 0.588 (0.463) & -0.031 (0.083) & \textbf{0.062 (0.050)} \\
        \hline 
         \multirow{2}{*}{3} & 1 & \textbf{-0.001 (0.061)} & 0.041 (0.037) & \textbf{0.029 (0.245)} & \textbf{0.157 (0.164)} & \textbf{-0.005 (0.222)} & 0.147 (0.135) & \textbf{0.003 (0.053)} & \textbf{0.034 (0.034)} \\
                            & 2 & -0.002 (0.056) & \textbf{0.038 (0.033)} & -0.212 (0.158) & 0.212 (0.139) & -0.010 (0.213) & \textbf{0.142 (0.131)} & -0.048 (0.031) & 0.049 (0.026) \\
        \hline 
         \multirow{2}{*}{4}   & 1 & \textbf{0.013 (0.158)} & \textbf{0.104 (0.098)} & \textbf{0.010 (0.273)} & \textbf{0.183 (0.167)} & \textbf{0.126 (0.271)} & \textbf{0.201 (0.181)} & \textbf{-0.007 (0.242)} & \textbf{0.163 (0.150)} \\
                              & 2 & 0.062 (0.146) & 0.105 (0.097) &  -0.205 (0.202) & 0.214 (0.163) & 0.565 (0.271) & 0.565 (0.263) & -0.173 (0.157) & 0.186 (0.113) \\
\hline
        \multirow{2}{*}{5}    & 1 & \textbf{0.001 (0.066)} & \textbf{0.045 (0.040)} &  \textbf{-0.007 (0.107)} & \textbf{0.071 (0.065)} & -0.001 (0.114) & \textbf{0.076 (0.069)} & \textbf{-0.007 (0.106)} & \textbf{0.071 (0.065)} \\
                              & 2 & \textbf{0.001 (0.066)} & \textbf{0.045 (0.040)} &  -0.225 (0.090) & 0.225 (0.090) & \textbf{0.000 (0.131)} & 0.089 (0.080) & -0.199 (0.070) & 0.199 (0.069) \\
         \hline 
         \multirow{2}{*}{6}   & 1 & \textbf{0.003 (0.131)} & \textbf{0.090 (0.079)} & \textbf{0.067 (0.433)} & 0.319 (0.259) & \textbf{0.104 (0.586)} & \textbf{0.342 (0.397)} & \textbf{0.018 (0.126)} & 0.075 (0.091) \\
                              & 2 & \textbf{0.003 (0.131)} & \textbf{0.090 (0.079)} &  -0.136 (0.352) & \textbf{0.267 (0.213)} & 0.534 (0.589) & 0.576 (0.461) & -0.031 (0.083) & \textbf{0.063 (0.049)} \\
        \hline 
         \multirow{2}{*}{7}   & 1 & \textbf{0.001 (0.061)} & 0.042 (0.037) & \textbf{0.048 (0.239)} & \textbf{0.157 (0.159)} & \textbf{0.000 (0.223)} & 0.149 (0.138) & \textbf{0.005 (0.052)} & \textbf{0.034 (0.033)} \\ 
                              & 2 & \textbf{0.001 (0.055)} & \textbf{0.038 (0.033)} & -0.202 (0.158) & 0.203 (0.140) & -0.001 (0.214) & \textbf{0.139 (0.133)} & -0.047 (0.031) & 0.047 (0.026) \\ 
        \hline 
         \multirow{2}{*}{8}   & 1 & -0.013 (0.149) & 0.102 (0.091) & \textbf{-0.002 (0.257)} & \textbf{0.174 (0.155)} & \textbf{-0.115 (0.275)} & \textbf{0.205 (0.181)} & \textbf{-0.019 (0.243)} & \textbf{0.160 (0.152)} \\
                              & 2 & \textbf{0.006 (0.143)} & \textbf{0.095 (0.087)} &  -0.204 (0.197) & 0.214 (0.154) & -0.541 (0.272) & 0.542 (0.258) & -0.185 (0.162) & 0.196 (0.113) \\
\hline \hline
\multirow{3}{*}{Scenario} & \multirow{3}{*}{Framework} & \multicolumn{8}{c|}{ $n=400$ Samples} \\
   \cline{3-10}
			 &  & \multicolumn{2}{c|}{$\hat{\theta}_0-\theta^*_0$} & \multicolumn{2}{c|}{$(\hat{A}-A^*)/A^*$} & \multicolumn{2}{c|}{$\hat{\eta}-\eta^*$} & \multicolumn{2}{c|}{$(\tau_{\text{Score}}-\tau^*_{\text{Score}})/n$} \\
   \cline{3-10}
   & & Bias & Abs. Err.& Bias & Abs. Err.& Bias & Abs. Err.& Bias & Abs. Err.  \\
   \hline
                \multirow{2}{*}{1}   & 1 & -0.001 (0.038) & 0.025 (0.023) & \textbf{-0.001 (0.070)} & \textbf{0.048 (0.042)} & -0.001 (0.069) & 0.047 (0.042) & \textbf{-0.006 (0.052)} & \textbf{0.036 (0.031)} \\
                                     & 2 & \textbf{0.000 (0.030)} & \textbf{0.020 (0.018)} & -0.228 (0.043) & 0.228 (0.043) & \textbf{0.000 (0.062)} & \textbf{0.042 (0.037)} & -0.203 (0.033) & 0.203 (0.033) \\
         \hline 
         \multirow{2}{*}{2}   & 1 & \textbf{-0.004 (0.069)} & 0.047 (0.042) & \textbf{-0.001 (0.232)} & \textbf{0.160 (0.140)} & \textbf{-0.121 (0.241)} & \textbf{0.180 (0.163)} & \textbf{-0.001 (0.057)} & \textbf{0.039 (0.035)} \\
                              & 2 & \textbf{-0.004 (0.068)} & \textbf{0.046 (0.041)} & -0.192 (0.186) & 0.198 (0.150) & -0.545 (0.240) & 0.545 (0.233) & -0.044 (0.038) & 0.046 (0.028) \\
        \hline 
         \multirow{2}{*}{3}    & 1 & -0.001 (0.030) & 0.020 (0.018) & \textbf{0.010 (0.110)} & \textbf{0.072 (0.069)} & \textbf{0.000 (0.111)} & 0.073 (0.068) & \textbf{0.001 (0.025)} & \textbf{0.017 (0.016)} \\ 
                               & 2 & \textbf{0.000 (0.028)} & \textbf{0.019 (0.017)} & -0.223 (0.079) & 0.223 (0.079) & -0.001 (0.108) & \textbf{0.070 (0.066)} & -0.050 (0.016) & 0.050 (0.015) \\ 
        \hline 
         \multirow{2}{*}{4}   & 1 & \textbf{0.013 (0.076)} & \textbf{0.053 (0.046)} & \textbf{-0.023 (0.129)} & \textbf{0.087 (0.081)} & \textbf{0.125 (0.136)} & \textbf{0.134 (0.109)} & \textbf{-0.025 (0.115)} & \textbf{0.080 (0.070)} \\
                              & 2 & 0.063 (0.071) & 0.069 (0.055) & -0.221 (0.100) & 0.221 (0.098) & 0.560 (0.134) & 0.560 (0.134) & -0.184 (0.076) & 0.184 (0.074) \\
\hline
        \multirow{2}{*}{5}  & 1 & \textbf{0.002 (0.034)} & \textbf{0.023 (0.020)} & \textbf{-0.009 (0.055)} & \textbf{0.039 (0.033)} & 0.003 (0.055) & \textbf{0.036 (0.033)} & \textbf{-0.008 (0.055)} & \textbf{0.039 (0.032)} \\
                            & 2 & \textbf{0.002 (0.034)} & \textbf{0.023 (0.020)} & -0.229 (0.046) & 0.229 (0.046) & \textbf{0.001 (0.061)} & 0.042 (0.037) & -0.203 (0.036) & 0.203 (0.036) \\
         \hline 
         \multirow{2}{*}{6}  & 1 & \textbf{-0.004 (0.068)} & \textbf{0.046 (0.041)} & \textbf{-0.013 (0.224)} & \textbf{0.154 (0.134)} & \textbf{0.121 (0.243)} & \textbf{0.182 (0.165)} & \textbf{-0.003 (0.056)} & \textbf{0.038 (0.035)} \\
                             & 2 & \textbf{-0.004 (0.068)} & \textbf{0.046 (0.041)} & -0.197 (0.183) & 0.203 (0.148) & 0.537 (0.246) & 0.537 (0.242) & -0.044 (0.038) & 0.046 (0.027) \\
        \hline 
         \multirow{2}{*}{7}  & 1 & -0.002 (0.030) & 0.020 (0.018) & \textbf{0.004 (0.108)} & \textbf{0.074 (0.065)} & -0.004 (0.114) & 0.075 (0.070) & \textbf{0.000 (0.025)} & \textbf{0.017 (0.015)} \\
                             & 2 & \textbf{-0.001 (0.028)} & \textbf{0.019 (0.017)} & -0.224 (0.078) & 0.224 (0.078) & \textbf{-0.001 (0.109)} & \textbf{0.072 (0.066)} & -0.050 (0.015) & 0.050 (0.015) \\
        \hline 
         \multirow{2}{*}{8}  & 1 & -0.017 (0.071) & \textbf{0.050 (0.044)} & \textbf{-0.027 (0.122)} & \textbf{0.087 (0.075)} & \textbf{-0.123 (0.131)} & \textbf{0.133 (0.101)} & \textbf{-0.031 (0.115)} & \textbf{0.081 (0.070)} \\
                             & 2 & \textbf{0.004 (0.069)} & 0.047 (0.041) & -0.215 (0.097) & 0.215 (0.095) & -0.541 (0.133) & 0.541 (0.133) & -0.191 (0.077) & 0.191 (0.074) \\
\hline \hline
\multirow{3}{*}{Scenario} & \multirow{3}{*}{Framework} & \multicolumn{8}{c|}{ $n=1,600$ Samples} \\
   \cline{3-10}
			 &  &  \multicolumn{2}{c|}{$\hat{\theta}_0-\theta^*_0$} & \multicolumn{2}{c|}{$(\hat{A}-A^*)/A^*$} & \multicolumn{2}{c|}{$\hat{\eta}-\eta^*$} & \multicolumn{2}{c|}{$(\tau_{\text{Score}}-\tau^*_{\text{Score}})/n$} \\
   \cline{3-10}
   & & Bias & Abs. Err.& Bias & Abs. Err.& Bias & Abs. Err.& Bias & Abs. Err.  \\
   \hline
                \multirow{2}{*}{1}   & 1 & \textbf{0.001 (0.019)} & 0.012 (0.011) & \textbf{-0.007 (0.035)} & \textbf{0.024 (0.021)} & 0.001 (0.034) & 0.023 (0.021) & \textbf{-0.007 (0.026)} & \textbf{0.018 (0.016)} \\
                                     & 2 & \textbf{0.001 (0.015)} & \textbf{0.010 (0.009)} & -0.230 (0.022) & 0.230 (0.022) & \textbf{0.000 (0.030)} & \textbf{0.020 (0.018)} & -0.204 (0.017) & 0.204 (0.017) \\
         \hline 
         \multirow{2}{*}{2}  & 1 & \textbf{0.000 (0.034)} & \textbf{0.023 (0.021)} & \textbf{-0.029 (0.114)} & \textbf{0.080 (0.071)} & \textbf{-0.123 (0.119)} & \textbf{0.127 (0.094)} & \textbf{-0.007 (0.028)} & \textbf{0.020 (0.017)} \\
                             & 2 & \textbf{0.000 (0.034)} & \textbf{0.023 (0.020)} & -0.212 (0.092) & 0.212 (0.090) & -0.543 (0.118) & 0.543 (0.118) & -0.047 (0.018) & 0.047 (0.018) \\
        \hline 
         \multirow{2}{*}{3}  & 1 & \textbf{0.000 (0.015)} & 0.010 (0.009) & \textbf{-0.007 (0.051)} & \textbf{0.035 (0.031)} & -0.002 (0.054) & 0.037 (0.033) & \textbf{-0.002 (0.012)} & \textbf{0.008 (0.007)} \\
                             & 2 & \textbf{0.000 (0.014)} & \textbf{0.009 (0.008)} & -0.232 (0.038) & 0.232 (0.038) & \textbf{-0.001 (0.053)} & \textbf{0.035 (0.032)} & -0.051 (0.008) & 0.051 (0.008) \\
        \hline 
         \multirow{2}{*}{4}  & 1 & \textbf{0.012 (0.038)} & \textbf{0.027 (0.024)} & \textbf{-0.037 (0.064)} & \textbf{0.051 (0.043)} & \textbf{0.127 (0.068)} & \textbf{0.127 (0.065)} & \textbf{-0.033 (0.058)} & \textbf{0.047 (0.038)} \\
                             & 2 & 0.060 (0.036) & 0.061 (0.033) & -0.230 (0.050) & 0.230 (0.050) & 0.559 (0.067) & 0.559 (0.067) & -0.188 (0.038) & 0.188 (0.038) \\
\hline
        \multirow{2}{*}{5}   & 1 & \textbf{0.000 (0.017)} & \textbf{0.012 (0.010)} & \textbf{-0.010 (0.027)} & \textbf{0.020 (0.017)} & \textbf{-0.001 (0.027)} & \textbf{0.019 (0.016)} & \textbf{-0.009 (0.027)} & \textbf{0.020 (0.017)} \\
                             & 2 & \textbf{0.000 (0.017)} & \textbf{0.012 (0.010)} & -0.231 (0.023) & \textbf{0.231 (0.023)} & \textbf{-0.001 (0.032)} & 0.021 (0.019) & -0.205 (0.017) & 0.205 (0.017) \\
         \hline 
         \multirow{2}{*}{6}  & 1 & \textbf{0.000 (0.034)} & \textbf{0.022 (0.020)} & \textbf{-0.026 (0.115)} & \textbf{0.079 (0.072)} & \textbf{0.125 (0.120)} & \textbf{0.128 (0.098)} & \textbf{-0.007 (0.028)} & \textbf{0.020 (0.017)} \\
                             & 2 & \textbf{0.000 (0.034)} & \textbf{0.022 (0.020)} & -0.209 (0.094) & 0.209 (0.093) & 0.546 (0.121) & 0.546 (0.121) & -0.047 (0.019) & 0.047 (0.018) \\
        \hline 
         \multirow{2}{*}{7}  & 1 & \textbf{0.000 (0.014)} & 0.010 (0.009) & \textbf{-0.007 (0.053)} & \textbf{0.036 (0.032)} & 0.003 (0.054) & 0.036 (0.033) & \textbf{-0.002 (0.012)} & \textbf{0.009 (0.007)} \\ 
                             & 2 & 0.001 (0.014) & \textbf{0.009 (0.008)} & -0.231 (0.039) & 0.231 (0.039) & \textbf{0.001 (0.052)} & \textbf{0.035 (0.032)} & -0.051 (0.008) & 0.051 (0.008) \\ 
        \hline 
         \multirow{2}{*}{8}  & 1 & -0.018 (0.036) & 0.028 (0.024) & \textbf{-0.033 (0.063)} & \textbf{0.048 (0.043)} & \textbf{-0.123 (0.065)} & \textbf{0.123 (0.062)} & \textbf{-0.032 (0.059)} & \textbf{0.046 (0.039)} \\ 
                             & 2 & \textbf{0.000 (0.035)} & \textbf{0.024 (0.021)} & -0.215 (0.050) & 0.215 (0.050) & -0.541 (0.065) & 0.541 (0.065) & -0.191 (0.039) & 0.191 (0.039) \\ 

\hline
\end{tabular}
}
\end{table*}

\clearpage

\newpage

\clearpage
\newpage

\begin{table*}[!h]
    \caption{Summary of each circadian transcriptomic data set before and after processing. ``GSCG'' denotes gold standard circadian gene.}
    \label{tab:data_sum}
      \centering
    \begin{tabular}{|c|c|c|c|c|c|c|c|c|c|}
    \hline
   \multirow{2}{*}{Data Set} & \multirow{2}{*}{Sample Population} & \multicolumn{2}{c|}{Number of Unique People} & \multicolumn{2}{c|}{Sample Size} & \multicolumn{2}{c|}{Number of Genes} & \multicolumn{2}{c|}{Number of GSCGs} \\
   \cline{3-10}
    & & Before & After & Before & After & Before & After & Before & After \\
        \hline 
         \multirow{3}{*}{Archer} & Control & 22 & 19 & 147 & 127 & 7,615 & 4,475 & 50 & 28 \\
         & Experimental & 21 & 19 & 139 & 131 & 7,615 & 4,599 & 50 & 31 \\
         & Combined & 43 & 38 & 286 & 258 & 7,615 & 3,689 & 50 & 23 \\
         \hline 
         Braun & Cohort & 11 & 11 & 153 & 153 & 7,615 & 7,615 & 50 & 50 \\
         \hline 
         \multirow{3}{*}{M\"{o}ller-Levet} & Control & 24 & 20 & 215 & 181 & 7,615 & 7,615 & 50 & 50 \\
         & Experimental & 24 & 20 & 212 & 174 & 7,615 & 7,615 & 50 & 50 \\
         & Combined & 48 & 40 & 427 & 355 & 7,615 & 7,615 & 50 & 50 \\
         \hline 
    \end{tabular}
\end{table*}

\clearpage
\newpage

\clearpage
\newpage

\begin{table*}[!h]
	\centering
	\caption{Empirical values of $\kappa^{(m)}(0)$ obtained from from Zeitgeber times of dim-light melatonin onset (DLMO times) recorded in 13 different studies (for $m=1,\ldots,8$). The smallest value in magnitude for each $\kappa^{(m)}(0)$ is denoted in bold.} 
	\label{tab:kappas}
 \resizebox{1.0\textwidth}{!}{
		\begin{tabular}{|l|c|c|c|c|c|c|c|c|}
			\hline
			Study & $\kappa^{(1)}(0)$ & $\kappa^{(2)}(0)$ & $\kappa^{(3)}(0)$ & $\kappa^{(4)}(0)$ & $\kappa^{(5)}(0)$ & $\kappa^{(6)}(0)$ & $\kappa^{(7)}(0)$ & $\kappa^{(8)}(0)$ \\ 
               \hline
\citealt{Agost}  & $\mathbf{1.841\times 10^{-1}}$ & $-8.342 \times 10^{-2}$ & $-2.004 \times 10^{-2}$ &  $4.498 \times 10^{-3}$ & $1.510\times 10^{-3}$ & $-1.893\times 10^{-4}$ & $-9.991 \times 10^{-5}$ & $5.490 \times 10^{-6}$  \\
\citealt{Akacem} & $1.037\times 10^{0}$ & $4.742\times 10^{-1}$ & $1.166 \times 10^{-1}$ & $1.248\times 10^{-2}$ & $-1.244 \times 10^{-3}$ & $-5.810\times 10^{-4}$ & $-4.026\times 10^{-5}$ & $1.491 \times 10^{-5}$  \\
\citealt{Burgess}  & $5.990\times 10^{-1}$ & $8.823\times 10^{-2}$ & $-2.319 \times 10^{-2}$ & $-9.139 \times 10^{-3}$ & $\mathbf{-1.693\times 10^{-4}}$ & $4.194\times 10^{-4}$ & $6.481\times 10^{-5}$ & $-1.022\times 10^{-5}$  \\
\citealt{Four} & $7.685 \times 10^{-1}$ & $2.190\times 10^{-1}$ & $\mathbf{1.055\times 10^{-2}}$ &$-1.046\times 10^{-2}$ & $-3.025\times 10^{-3}$ & $-1.950\times 10^{-4}$ & $8.788 \times 10^{-5}$ & $2.814 \times 10^{-5}$ \\
\citealt{Kantermann2015} & $9.647 \times 10^{-1}$ & $4.414 \times 10^{-1}$ & $1.259\times 10^{-1}$ & $2.469\times 10^{-2}$ & $3.455\times 10^{-3}$ & $3.465 \times 10^{-4}$ & $2.484\times 10^{-5}$ & $\mathbf{1.434\times 10^{-6}}$  \\
\citealt{kawinska}  & $1.142\times 10^{0}$ & $6.361 \times 10^{-1}$ & $2.302 \times 10^{-1}$ & $6.085 \times 10^{-2}$  &  $1.252 \times 10^{-2}$ & $2.088\times 10^{-3}$ & $2.897 \times 10^{-4}$ & $3.406 \times 10^{-5}$ \\
\citealt{kitamura} & $8.005 \times 10^{-1}$ & $2.614 \times 10^{-1}$ & $3.617 \times 10^{-2}$ & $-1.318 \times 10^{-3}$ & $-1.068\times 10^{-3}$ & $\mathbf{-4.847\times 10^{-5}}$ & $2.785 \times 10^{-5}$ & $3.312 \times 10^{-6}$ \\
\citealt{Martin} & $5.770\times 10^{-1}$ & $4.693 \times 10^{-2}$ & $-4.640 \times 10^{-2}$ & $-1.264 \times 10^{-2}$ & $1.338 \times 10^{-3}$ & $1.096\times 10^{-3}$ & $7.964 \times 10^{-5}$ & $-5.902 \times 10^{-5}$ \\
\citealt{mcglashan} & $7.358\times 10^{-1}$ & $2.304\times 10^{-1}$ & $3.485 \times 10^{-2}$ & $\mathbf{2.849\times 10^{-4}}$ & $-1.050\times 10^{-3}$ & $-2.313 \times 10^{-4}$ & $-1.658\times 10^{-5}$ & $3.571 \times 10^{-6}$ \\
\citealt{mcmahon} & $7.508\times 10^{-1}$ & $1.655 \times 10^{-1}$ & $-3.463 \times 10^{-2}$ & $-2.753 \times 10^{-2}$ & $-4.675 \times 10^{-3}$ & $1.096 \times 10^{-3}$ & $6.997 \times 10^{-4}$ & $1.050 \times 10^{-4}$ \\
\citealt{roemer} & $4.960 \times 10^{-1}$ & $4.015\times 10^{-2}$ & $-2.832 \times 10^{-2}$ & $-8.674\times 10^{-3}$ & $-3.323\times 10^{-4}$ & $3.581 \times 10^{-4}$ & $9.786 \times 10^{-5}$ & $6.355 \times 10^{-6}$ \\
\citealt{Ruiz2020} & $3.180\times 10^{-1}$ & $\mathbf{-2.335 \times 10^{-2}}$ & $-1.422\times 10^{-2}$ & $8.794 \times 10^{-4}$ & $3.509 \times 10^{-4}$ & $-5.248\times 10^{-5}$ & $\mathbf{-1.707\times 10^{-6}}$ & $2.189\times 10^{-6}$ \\
\citealt{Wittenbrink2018} & $6.847 \times 10^{-1}$ & $1.083 \times 10^{-1}$ & $-3.238 \times 10^{-2}$ & $-9.932\times 10^{-3}$ & $1.582\times 10^{-3}$ & $7.000 \times 10^{-4}$ & $-8.548\times 10^{-5}$ & $-4.790 \times 10^{-5}$ \\
Combined & $7.091\times 10^{-1}$ & $1.455\times 10^{-1}$ & $-2.450 \times 10^{-2}$ & $-1.671 \times 10^{-2}$ & $-1.850 \times 10^{-3}$ & $6.575 \times 10^{-4}$ & $2.297 \times 10^{-4}$ & $6.273 \times 10^{-6}$ \\
\hline
\end{tabular}}
\end{table*}

\clearpage
\newpage

\begin{table*}[!h]
	\caption{Comparison of both frameworks using every gene available. Framework 1 uses the corrected score function from Theorem \ref{thm1} for obtaining each quantity. Framework 2 uses a score function that does not account for measurement error. The regression parameter estimate $\beta$ is listed alongside the coefficient of determination ($R^2$), where the latter is in parentheses. Bold values indicate a value of $\beta$ closer to one, which signifies that the quantities obtained from a framework with mismeasured covariate data are closer to the quantities obtained with correctly measured data.} \label{tab:app1}
 \centering
		\begin{tabular}{|l|c|c|c|c|c|c|c|c|c|}
			\hline
   \multirow{1}{*}{Sample Population} & \multirow{1}{*}{Framework}  &  \multicolumn{1}{c|}{$A^*$} & \multicolumn{1}{c|}{$\eta^*$} & \multicolumn{1}{c|}{$\tau^*_{\text{Score}}/n$} \\
   \hline
         \multirow{2}{*}{Archer (Control)} & 1 & 0.965 (0.988) & \textbf{0.925 (0.937)} & 0.910 (0.964) \\ 
                                            & 2       &  \textbf{1.016 (0.987)} & 0.873 (0.893) & \textbf{0.995 (0.963)} \\ 
         \hline 
         \multirow{2}{*}{Archer (Experimental)} & 1 & 0.963 (0.926) & \textbf{0.881 (0.919)} & 0.877 (0.807) \\ 
                                               & 2           & \textbf{1.010 (0.927)} & 0.874 (0.915) & \textbf{0.955 (0.808)} \\ 
         \hline 
         \multirow{2}{*}{Archer (Combined)}& 1 & \textbf{1.028 (0.962)} & \textbf{0.896 (0.920)} & \textbf{1.046 (0.905)} \\ 
                                                    & 2  & 1.081 (0.961) & 0.867 (0.893) & 1.138 (0.902) \\ 
         \hline 
         \multirow{2}{*}{Braun} & 1 & \textbf{0.971 (0.991)} & \textbf{0.954 (0.952)} & \textbf{1.012 (0.983)} \\ 
                                & 2           & 1.030 (0.991) & 0.913 (0.918) & 1.115 (0.983) \\ 
         \hline 
         \multirow{2}{*}{M\"{o}ller-Levet (Control)} & 1 & \textbf{1.105 (0.970)} & \textbf{0.926 (0.930)} & \textbf{1.173 (0.948)} \\ 
                                                 & 2           & 1.148 (0.971) & 0.912 (0.938) & 1.266 (0.950) \\ 
         \hline 
         \multirow{2}{*}{M\"{o}ller-Levet (Experimental)} & 1 &  \textbf{1.037 (0.985)} & \textbf{0.967 (0.915)} & \textbf{1.041 (0.973)} \\ 
                                                      & 2           &  1.076 (0.985) & \textbf{0.967 (0.944)} & 1.122 (0.973) \\ 
         \hline 
         \multirow{2}{*}{M\"{o}ller-Levet (Combined)} & 1 & \textbf{1.087 (0.982)} & \textbf{0.956 (0.928)} & \textbf{1.137 (0.967)} \\ 
                                                  & 2           &  1.128 (0.982) & 0.944 (0.945) & 1.225 (0.968) \\ 
         \hline 
\end{tabular} 
\end{table*}

\clearpage
\newpage

\begin{table*}[!h]
	\caption{Comparison of both frameworks using every gold standard circadian gene available. Framework 1 uses the corrected score function from Theorem \ref{thm1} for obtaining each quantity. Framework 2 uses a score function that does not account for measurement error. The regression parameter estimate $\beta$ is listed alongside the coefficient of determination ($R^2$), where the latter is in parentheses. Bold values indicate a value of $\beta$ closer to one, which signifies that the quantities obtained from a framework with mismeasured covariate data are closer to the quantities obtained with correctly measured data.} \label{tab:app2}
 \centering
		\begin{tabular}{|l|c|c|c|c|c|c|c|c|c|c|c|}
			\hline
   \multirow{1}{*}{Data Set} & \multirow{1}{*}{Framework} & \multicolumn{1}{c|}{$A^*$} & \multicolumn{1}{c|}{$\eta^*$} & \multicolumn{1}{c|}{$\tau^*_{\text{Score}}/n$} \\
   \hline
         \multirow{2}{*}{Archer (Control)} & 1 & 0.959 (0.991) & \textbf{0.872 (0.943)} & \textbf{0.982 (0.986)} \\ 
                                            & 2 &\textbf{1.008 (0.991)} & 0.835 (0.907) & 1.068 (0.986) \\ 
         \hline 
         \multirow{2}{*}{Archer (Experimental)} & 1 & \textbf{1.068 (0.954)} & 0.998 (0.977) & \textbf{1.342 (0.870)} \\ 
                                                & 2 & 1.115 (0.952) & \textbf{0.999 (0.969)} & 1.453 (0.864) \\ 
         \hline 
         \multirow{2}{*}{Archer (Combined)} & 1 & \textbf{1.039 (0.961)} & \textbf{1.029 (0.921)} & \textbf{1.001 (0.945)} \\ 
                                            & 2 & 1.087 (0.962) & 1.032 (0.896) & 1.085 (0.945) \\ 
         \hline 
         \multirow{2}{*}{Braun} & 1 &  0.945 (0.994) & \textbf{0.940 (0.938)} & 0.912 (0.991) \\ 
                                & 2          &  \textbf{0.996 (0.993)} & 0.907 (0.902) & \textbf{0.996 (0.990)} \\ 
         \hline 
         \multirow{2}{*}{M\"{o}ller-Levet (Control)} & 1 &  \textbf{1.046 (0.969)} & \textbf{0.967 (0.926)} & \textbf{1.086 (0.963)} \\ 
                                                 & 2           &  1.093 (0.970) & 0.949 (0.922) & 1.182 (0.965) \\ 
         \hline 
         \multirow{2}{*}{M\"{o}ller-Levet (Experimental)} & 1 & \textbf{1.053 (0.988)} & \textbf{0.957 (0.929)} & \textbf{1.112 (0.987)} \\ 
                                                     & 2           &  1.096 (0.987) & \textbf{0.957 (0.948)} & 1.205 (0.985) \\ 
         \hline 
         \multirow{2}{*}{M\"{o}ller-Levet (Combined)} & 1 & \textbf{1.061 (0.985)} & 1.010 (0.952) & \textbf{1.109 (0.986)} \\ 
                                                  & 2     &  1.106 (0.986) & \textbf{0.998 (0.964)} & 1.204 (0.985) \\ 
         \hline 
\end{tabular} 
\end{table*}

\newpage

\appendix

\section{Theoretical Results} \label{app:A}
\subsection{Supporting lemmas}

\begin{lemma}[Page 26, \citealt{gelfand1964}] \label{lem:1.1}
Suppose $h(Z)$ is a function that is $m$-times continuously differentiable with a univariate argument $Z$, and define $W$ to be a univariate constant. Then $\int h(Z)\delta^{(m)}(Z-W)dZ = (-1)^mh^{(m)}(W)$.
\end{lemma}

\begin{lemma} \label{lem:1.2}
The Fourier transform for the $m$-th order derivative of a Dirac delta function $\delta^{(m)}(Z-W)$ is given by $(-it)^m\exp(itW)$.
\begin{proof}
By application of Lemma \ref{lem:1.1}, $\int \exp(itZ)\delta^{(m)}(Z-W)dZ = (-it)^m\exp(itW)$.
\end{proof}
\end{lemma}

\begin{lemma} \label{lem:1.3}
The $m$-th order derivative of the Dirac delta function $\delta^{(m)}(Z-W) = \frac{1}{2\pi}\int (-it)^m\exp\{it(W-Z)\} dt$.
\begin{proof}
First, note that the inverse Fourier transform is defined as
\begin{align} \label{def:inv}
    h(Z) = \frac{1}{2\pi}\int \phi_Z(t)\exp(-itZ)dZ,
\end{align}
where $\phi_Z(t)$ is the Fourier transform of $h(Z)$ with respect to its argument $Z$, or
\begin{align}
\phi_Z(t) = \int h(Z)\exp(itZ)dZ. \label{eq:four}
\end{align}
From Lemma \ref{lem:1.2}, we can plug $\phi_Z(t) = (-it)^m\exp(itW)$ into (\ref{def:inv}) to obtain $\delta^{(m)}(Z-W) = \frac{1}{2\pi}\int (-it)^m\exp\{it(W-Z)\} dt$. 
\end{proof}
\end{lemma}

\subsection{Proof of Theorem \ref{thm1}}
The expected score function $\mathbb{E}\{g(Y, X^{*}; \theta)\}$ is expressed as
\begin{align}
\mathbb{E}\left\{g(Y, X^{*}; \theta)\right\}=& \int\rho(Y)\int \rho(X^{*} \mid Y)g(Y, X^{*}; \theta)dX^{*}dY \nonumber \\
 =&\frac{1}{2\pi}\int\rho(Y)\int g(Y, X^{*}; \theta)\int \phi_{X^{*} \mid Y}(t)\exp(-itX^{*})dtdX^{*}dY, \label{eq:2.1}
\end{align}
where (\ref{eq:2.1}) is by definition of the basic inversion of the Fourier transform for a probability density function from (\ref{def:inv}). Now, recall that the Fourier transform for a sum of independent random variables is equal to the product of Fourier transforms for each random variable. It follows that the Fourier transform of the conditional density $\phi_{X^{\dagger}\mid Y}(t)=\phi_{X^*+\xi \mid Y}(t)=\phi_{X^{*}\mid Y}(t)\phi_{\xi\mid Y}(t)$, as $\xi$ is independent of $X^*$ and $Y$. Further, because $\xi$ and $Y$ are independent, $\phi_{\xi \mid Y}(t) = \phi_{\xi}(t)$. Now, define $\kappa(t) = 1/\mathbb{E}\{\exp(t\xi)\}$, which implies that $\kappa(it) = 1/\phi_{\xi}(t)$ and yields $\phi_{X^{*}\mid Y}(t)=\phi_{X^{\dagger}\mid Y}(t)\kappa(it)$. The expected score function becomes
\begin{align}
    \mathbb{E}\left\{g(Y, X^{*}; \theta)\right\} =&\frac{1}{2\pi}\int \rho(Y)\int g(Y, X^{*}; \theta)\int \phi_{X^{\dagger}\mid Y}(t)\kappa(it)\exp(-itX^{*})dtdX^{*}dY \nonumber \\
    =& \frac{1}{2\pi}\int \rho(Y)\int g(Y, X^{*}; \theta) \nonumber \\
    &\quad \times \int\phi_{X^{\dagger}\mid Y}(t)\left\{\sum_{m=0}^{\infty}\frac{\kappa^{(m)}(0)(it)^m}{m!}\right\}\exp(-itX^{*}) dtdX^{*}dY \label{eq2.2} \\
    =& \frac{1}{2\pi}\sum_{m=0}^{\infty}\frac{\kappa^{(m)}(0)}{m!}\int \rho(Y) \int g(Y, X^{*}; \theta) \nonumber \\
    & \quad \times \int \phi_{X^{\dagger}\mid Y}(t)(it)^m\exp(-itX^{*})dt dX^{*}dY, \nonumber
\end{align}
where (\ref{eq2.2}) uses a Taylor expansion on $\kappa(it)$ centered at $0$ under the assumption that $\phi_{\xi}(t)$ is smooth and non-zero for all arguments $t$. Substituting $\phi_{X^{\dagger}\mid Y}(t)$ with its definition from (\ref{eq:four}), it follows that
\begin{align}
    \mathbb{E}\left\{g(Y, X^{*}; \theta)\right\}
    =& \frac{1}{2\pi}\sum_{m=0}^{\infty}\frac{\kappa^{(m)}(0)}{m!}\int \rho(Y) \int g(Y, X^{*}; \theta) \int \rho(X^{\dagger}\mid Y) \nonumber \\
    & \quad \times \int (it)^m\exp\left\{it(X^{\dagger}-X^{*})\right\}dtdX^{\dagger} dX^{*}dY \nonumber \\
    =& \frac{1}{2\pi}\sum_{m=0}^{\infty}\frac{\kappa^{(m)}(0)}{m!}\int \rho(Y)\int\rho(X^{\dagger}\mid Y)  \nonumber \\
    & \quad \times \int g(Y, X^{*}; \theta)\int (it)^m\exp\left\{it(X^{\dagger}-X^{*})\right\}dt dX^{*}dX^{\dagger}dY \nonumber \\
    =& \sum_{m=0}^{\infty}\frac{\kappa^{(m)}(0)}{m!}\int \rho(Y)\int \rho(X^{\dagger}\mid Y) \nonumber \\
    & \quad \times \int g(Y, X^{*}; \theta)(-1)^m\delta^{(m)}(X^{*}-X^{\dagger})dX^{*}dX^{\dagger}dY  \label{eq2.4} \\
    =& \sum_{m=0}^{\infty}\frac{\kappa^{(m)}(0)}{m!} \int \rho(Y) \int \rho(X^{\dagger} \mid Y)\left\{\frac{d^mg(Y, X^{\dagger}; \theta)}{d(X^{\dagger})^m}\right\}dX^{\dagger}dY \label{eq2.5} \\
    =&\sum_{m=0}^{\infty}\frac{\kappa^{(m)}(0)}{m!}\mathbb{E}\left\{\frac{d^mg(Y, X^{\dagger}; \theta)}{d(X^{\dagger})^m} \right\}. \nonumber
\end{align}
Here, (\ref{eq2.4}) is due to Lemma \ref{lem:1.3}, and (\ref{eq2.5}) is due to Lemma \ref{lem:1.1}. 

\end{document}